\documentclass[aps, prb, reprint, superscriptaddress, floatfix]{revtex4-2}

\usepackage{xcolor}
\usepackage{booktabs}
\usepackage{graphicx}% Include figure files
\usepackage{dcolumn}% Align table columns on decimal point
\usepackage{amssymb}
\usepackage{siunitx}
\usepackage{amsmath}
\usepackage[T1]{fontenc}
\usepackage[utf8]{inputenc}
\usepackage{hyperref}% add hypertext capabilities
\hypersetup{colorlinks=true, citecolor=blue, linkcolor=blue, urlcolor=blue}
\newcommand{\br}{\textbf{r}}
\newcommand{\bR}{\textbf{R}}
\newcolumntype{d}[1]{D{.}{.}{#1}}

\begin{document}

\title{Exotic Harmonium Model: Exploring Correlation Effects of Attractive Coulomb Interaction}

\author{Nahid Sadat Riyahi}
\affiliation{Department of Physical and Computational Chemistry, Shahid Beheshti University, Evin, Tehran 19839-69411, Iran}
\author{Mohammad Goli}
\email{m{\_}goli@ipm.ir}
\affiliation{School of Quantum Physics and Matter, Institute for Research in Fundamental Sciences (IPM), Tehran 19538-33511, Iran}
\author{Shant Shahbazian}
\email{sh{\_}shahbazian@sbu.ac.ir}
\affiliation{Department of Physics, Shahid Beheshti University, Evin, Tehran 19839-69411, Iran}

\date{\today}

\begin{abstract}
Simple few-body systems often serve as theoretical laboratories across various branches of theoretical physics. A prominent example is the two-electron Harmonium model, which has been widely studied over the past three decades to gain insights into the nature of the electron-electron correlations in many-electron quantum systems. Building on our previous work [Phys. Rev. B 108, 245155 (2023)], we introduce an analogous model consisting of an electron and a positively charged particle (PCP) with variable mass, interacting via Coulomb forces while confined by external harmonic potentials. Termed the \textit{exotic Harmonium} model, this provides insights into the electron-PCP correlations, a cornerstone of the emerging field of the \textit{ab initio} study of multi-component many-body quantum systems. Through a systematic exploration of the parameter space and numerical solutions of the corresponding Schrödinger equation, we identify two extreme regimes: the atom-like and the particle-in-trap-like behavior. The electron-PCP correlation dominates in the atom-like regime, significantly influencing physical observables, while its role diminishes in the particle-in-trap-like limit. Between these two extremes lies a complex intermediate regime that challenges qualitative interpretation. Overall, the exotic Harmonium model offers a powerful framework to unravel the electron-PCP correlations across diverse systems, spanning particles of varying masses and conditions, from ambient to high-pressure environments.                            
\end{abstract}


\maketitle

\section{Introduction}%
The concept of electron-electron correlation is a central tenet of the modern electronic structure theory and a primary focus in the physics of the many-electron systems \cite{helgaker_molecular_2000,martin_electronic_2008,fulde_correlated_2012,coleman_introduction_2016}. However, there are many intriguing systems and phenomena in nature where, in addition to electrons, other types of quantum particles play a crucial role in the underlying physics. Notable examples include the proton-coupled electron transfer \cite{weinberg_proton-coupled_2012,hammes-schiffer_proton-coupled_2015}, high-pressure superconducting hydrogen-rich materials \cite{pickard_superconducting_2020,pickett_colloquium_2023}, and the exotic condensed phases of hydrogen under extreme pressures \cite{kitamura_quantum_2000,silvera_phases_2021}. In these examples, protons must preferably be treated as quantum particles rather than clamped point charges. Similarly, positrons and positively charged muons should also be treated as quantum particles when they interact at low energies with molecules and condensed phases, forming bound states upon attachment \cite{puska_theory_1994,gribakin_positron-molecule_2010,mills_physics_2011,tuomisto_defect_2013,bass_colloquium_2023,storchak_quantum_1998,nagamine_introductory_2003,blundell_muon-spin_2004,hillier_muon_2022}. 

In all these systems and phenomena, the fundamental “multicomponent” (MC) Hamiltonian (see Eq. \ref{eq:1}) includes the kinetic energy operators of both electrons and protons/muons/positrons. Hereafter, the latter particles are collectively referred to as positively charged particles (PCPs). The \textit{ab initio} solution of the eigenvalue problem of the MC Hamiltonian has been studied systematically only in the last two decades \cite{nakai_nuclear_2007,ishimoto_review_2009,reyes_any_2019,hammes-schiffer_nuclearelectronic_2021}, and has been recently recognized as a distinct subfield of quantum chemistry, termed as the MC quantum chemistry \cite{pavosevic_multicomponent_2020}. What distinguishes the MC quantum chemistry from the brute-force MC \textit{ab initio} procedures, which aim to find the variationally optimized explicitly correlated MC wavefunctions \cite{bubin_bornoppenheimer_2013,mitroy_theory_2013}, is its foundational premise of assigning spin-orbitals to both electrons and PCPs. This represents a significant advantage, as the orbital-based \textit{ab initio} methods are computationally cost-effective and can be readily extended to many-atom systems \cite{pavosevic_multicomponent_2020}. 

Consequently, the resulting MC wavefunctions are typically straightforward to interpret in chemical terms using modern partitioning algorithms \cite{shahbazian_mc-qtaim_2022}. Initially, there were disagreements regarding the optimal formulation of MC quantum chemical methods, even at the level of selecting the appropriate form of the MC Hamiltonian \cite{ishimoto_review_2009,tachikawa_extension_1998,webb_multiconfigurational_2002,nakai_simultaneous_2002,nakai_nuclear_2007,reyes_any_2019}. However, a consensus now appears to have emerged, favoring an approach where electrons and certain particles in the system are treated simultaneously as quantum particles, while the rest are modeled as clamped point charges \cite{hammes-schiffer_nuclearelectronic_2021,pavosevic_multicomponent_2020}. This approach eliminates the need for a complex procedure of removing the center-of-mass motion from the MC Hamiltonian \cite{bochevarov_electron_2004,sutcliffe_molecular_2005}, while retaining the advantageous feature of the adiabatic separability. Additionally, it avoids the unnecessary dynamical coupling of all particles of the system simultaneously \cite{essen_physics_1977}. Let us briefly review the fundamental structure of MC quantum chemistry and its broader implications. 

The basic Hamiltonian of a general many-body Coulombic molecular or condensed-phase system is as follows (throughout the paper, all quantities are given in atomic units except otherwise stated):
\begin{align}\label{eq:1}
    {{\hat{H}}_\mathrm{MC}}=&-\sum\limits_{k}^{M}{\sum\limits_{i}^{{{N}_{k}}}{\frac{\nabla _{i,k}^{2}}{2{{m}_{k}}}}}+\sum\limits_{k}^{M}{\sum\limits_{l>k}^{M}{\sum\limits_{i}^{{{N}_{k}}}{\sum\limits_{j}^{{{N}_{k}}}{\frac{{{q}_{k}}{{q}_{l}}}{\left| {{{\br}}_{i,k}}-{{{\br}}_{j,l}} \right|}}}}} \nonumber \\ 
    &+\sum\limits_{k}^{M}{\sum\limits_{i}^{{{N}_{k}}}{\sum\limits_{j>i}^{{{N}_{k}}}{\frac{q_{k}^{2}}{\left| {{{\br}}_{i,k}}-{{{\br}}_{j,k}} \right|}}}} \nonumber \\
    &+\sum\limits_{k}^{M}{\sum\limits_{i}^{{{N}_{k}}}{\sum\limits_{\alpha }^{Q}{\frac{{{Z}_{\alpha }}{{q}_{k}}}{\left| {{{\bR}}_{\alpha }}-{{{\br}}_{i,k}} \right|}}}}.
\end{align}

This Hamiltonian describes a system composed of $M$ types of quantum particles, each type containing $N_k$ particles with the mass $m_k$ and the charge $q_k$. These particles interact with each other and an external Coulomb field generated by $Q$ clamped particles, each carrying $Z_\alpha$ units of the electron’s charge. It represents a diverse set of systems, such as positronic, muonic and protonic systems, by assigning appropriate values to the masses and charges. Similar to the case of conventional “purely” electronic structure theory \cite{helgaker_molecular_2000}, the hierarchical structure of many-body MC quantum chemistry begins with uncorrelated mean-field theory, specifically the MC-Hartree-Fock method (MC-HF) \cite{pavosevic_multicomponent_2020}. Subsequently, more accurate wavefunctions and energies are derived through various post-MC-HF procedures, such as MC many-body perturbation theory, MC coupled-cluster and MC configuration-interaction methods, progressively converging toward the exact solution \cite{pavosevic_multicomponent_2020}. This hierarchical structure performs well in recovering the electron-electron correlation effects on various observables \cite{helgaker_molecular_2000}, achieving the “chemical accuracy”, i.e., ~ 1 {kcal.mol}$^{-1}$ accuracy for relative energies. However, its inefficiency in capturing electron-PCP correlation effects has been well-documented, pointing to the slow convergence of the post-MC-HF methods \cite{pak_electron-proton_2004,chakraborty_inclusion_2008,chakraborty_erratum_2011,ko_alternative_2011,swalina_analysis_2012,brorsen_nuclear-electronic_2015}. This limitation is partly understandable, as the electron-PCP interactions are attractive, whereas the success of the conventional post-HF methods has been primarily demonstrated for repulsive electron-electron interactions \cite{helgaker_molecular_2000}. The single-determinant HF wavefunction typically serves well as a reliable starting point in the hierarchical structure of post-HF methods, as it incorporates exchange correlation through the antisymmetric nature of the determinant. However, electrons and PCPs represent “distinguishable” groups of particles, lacking any exchange correlation. Consequently, the MC-HF wavefunction, which is a Hartree-like product of Slater determinants (for fermions) and/or permanents (for bosons) for each group of distinguishable particles, is intrinsically a crude starting point for recovering electron-PCP correlation \cite{cassam-chenai_decoupling_2015,cassam-chenai_quantum_2017}. This dilemma currently lacks a proper solution, and the “holy grail” of the MC quantum chemistry lies in finding an appropriate ansatz for efficiently recovering electron-PCP correlation. In principle, this could be achieved either by starting from a new initial point, distinct from the MC-HF wavefunction, or by developing a novel post-MC-HF methodology with a high convergence rate. 

As an alternative, the \textit{ab initio} MC density functional theory (MC-DFT) has been developed \cite{pavosevic_multicomponent_2020}. However, the challenge of accurately recovering electron-PCP correlation now manifests in the need to develop efficient electron-PCP correlation functionals \cite{chakraborty_development_2008,chakraborty_erratum_2011-1,kuriplach_improved_2014,zubiaga_full-correlation_2014,yang_development_2017,brorsen_alternative_2018,tao_multicomponent_2019,goli_two-component_2022}. A recent study by our research group provides an ambivalent assessment of some of the more widely used electron-proton/muon correlation functionals, highlighting inherent deficiencies in their functional design strategies \cite{riyahi_quantifying_2023}. In addition to electron-PCP correlation, the PCP-PCP correlation also arises in multi-PCP quantum systems, though this topic is beyond the scope of the present paper \cite{gidopoulos_kohn-sham_1998}. Finally, it is important to emphasize that the inter-dependence of various types of correlations in MC quantum systems remains a critical subject that warrants further investigations in future studies \cite{brorsen_alternative_2018,sirjoosingh_multicomponent_2012,udagawa_electron-nucleus_2014}.

To gain a clearer understanding of the nature of electron-proton/muon correlation, we introduced a two-particle model \cite{riyahi_quantifying_2023,goli_comment_2023}. This model consists of an electron and a proton/muon in a double harmonic trap. Despite its simplicity, the model proved to be highly effective in evaluating the quality of electron-proton/muon correlation functionals, uncovering certain shortcomings that had not been previously identified. A key advantage of this model is its freedom from interfering effects of other types of correlations, such as electron-electron interactions, which are inevitably present in real molecules. In this regard, the model serves as a “clean laboratory” to isolate and focus exclusively on the electron-PCP correlation. 

In the present study, we focus on exploring the model by systematically varying its parameters, specifically, the mass of PCP and the strength of the external field across a wide range of values. This methodology enables a comprehensive survey of the role of electron-PCP correlation in vastly different environments. In contrast, much of our current understanding of the role and effects of electron-PCP correlation in real molecules or condensed phases stems from case studies on species containing PCPs with certain masses, predominantly proton, typically examined under ambient pressure conditions. In the absence of robust theoretical arguments regarding the explicit mass-dependence of electron-PCP correlation effects, extrapolating findings from a case study involving one type of PCP to species with different PCPs is inherently unreliable. Additionally, under high hydrostatic pressures, atoms in a solid move closer together, creating novel chemical environments for electrons and PCPs that differ significantly from those under ambient pressure conditions \cite{hemley_revealing_1998,mcmillan_new_2002,grochala_chemical_2007,pickett_quest_2019,tse_chemical_2020}. In such scenarios, electron-PCP correlation may manifest in new ways not observed in matter at ambient pressures. We hope that this model opens a new avenue for the much-needed “unified” study of electron-PCP correlation effects across a broad spectrum of conditions, a domain that remains largely unexplored in the relevant literature. 

\section{Theory}

The Helium atom is arguably the simplest natural electronic system in which electron-electron correlation manifests itself, aside from artificially designed systems such as Helium-like geonium atom \cite{m_martins_heliumlike_2001}, or other artificial two-electron atoms \cite{ashoori_electrons_1996,yannouleas_collective_2000}. Despite its simplicity, the corresponding electronic Schrödinger equation for Helium remains unsolved analytically \cite{bartlett_normal_1935,withers_schrodinger_1984,withers_further_2011,toli_schrodinger_2019}, although highly accurate and complex wavefunctions have been developed as approximate solutions \cite{brown_analytic_1967,brown_integral_1967,brown_analytic_1967-1,hylleraas_schrodinger_1964,schwartz_experiment_2006,tanner_theory_2000,drake_high_2006,aznabaev_nonrelativistic_2018}. 

To study electron-electron correlation effects in a computationally less demanding manner, Kestner and Sinanoglu introduced a simplified model of the Helium atom, known as Harmonium or Hooke’s atom \cite{kestner_study_1962}. Harmonium was proposed as a hypothetical two-electron system in which the Coulombic attraction terms between electrons and the nucleus in the Hamiltonian of the Helium atom are replaced with harmonic potentials. The resulting Hamiltonian is as follows:

\begin{equation}
    {{\hat{H}}_\mathrm{Har}}=-\frac{1}{2}\nabla _{1}^{2}-\frac{1}{2}\nabla _{2}^{2}+\frac{1}{2}{{\omega }^{2}}\left( r_{1}^{2}+r_{2}^{2} \right)+\frac{1}{\left| {{{\br}}_{1}}-{{{\br}}_{2}} \right|}.
\end{equation}

Through a transformation to the center-of-mass, $\bR=\left( {{{\br}}_{1}}+{{{\br}}_{2}} \right)/2$, and relative, $\br={{\br}_{2}}-{{\br}_{1}}$, coordinates, the Hamiltonian of Harmonium, unlike that of the Helium atom, becomes separable into two uncoupled single-pseudo-particle Hamiltonians \cite{kestner_study_1962}. The center-of-mass Hamiltonian reduces to a harmonic oscillator model, for which the eigenvalue problem is analytically solvable \cite{sakurai_modern_2010}. The relative-coordinate Hamiltonian, however, contains both harmonic and Coulomb potentials. While Kestner and Sinanoglu did not originally recognize this, its eigenvalue problem belongs to the class of quasi-exactly solvable models \cite{kais_dimensional_1989,taut_two_1993}. As a result, the eigenvalue problem is analytically solvable only for specific values of the frequency of oscillation, $\omega$, of the Harmonium model \cite{ushveridze_quasi-exactly_1994}. Interestingly, introducing an additional linear term to the electron-electron repulsive Coulomb potential in Harmonium, despite being a more radical deviation from the Hamiltonian of the Helium atom, results in an exactly solvable model \cite{samanta_correlation_1990,ghosh_study_1991,turbiner_two_1994}. Since analytical solutions of the Harmonium are readily available, the role of electron-electron correlation in determining various properties of the system can be studied in detail \cite{santos_calculo_1968,tuan_double_1969,white_perturbation_1970,benson_perturbation_1970,king_electron_1996,zhu_size_1997,lamouche_two_1998,cioslowski_ground_2000,henderson_electron_2001,romera_electron-pair_2002,cyrnek_energy_2003,oneill_wave_2003,mandal_two_2003,amovilli_approximate_2003,karwowski_harmonium_2004,gill_electron_2005,katriel_effect_2005,akbari_momentum_2007,ragot_comments_2008,loos_correlation_2009,matito_properties_2010,ebrahimi-fard_harmonium_2012,karwowski_biconfluent_2014,nagy_information-theoretic_2015,cioslowski_harmonium_2017,galiautdinov_ground_2018,cioslowski_natural_2018,rusin_pauli_2021}. In particular, the model has been widely used to calibrate the accuracy of the electronic exchange-correlation functionals and served as a test bed for applying and evaluating the effectiveness of various theoretical ideas developed within the framework of the electronic DFT \cite{laufer_test_1986,hall_comparison_1989,samanta_density-functional_1991,samanta_study_1991,kais_density_1993,filippi_comparison_1994,burke_semilocal_1995,huang_local_1997,taut_two_1998,qian_physics_1998,march_differential_1998,lam_virial_1998,hessler_several_1999,hessler_erratum_1999,ivanov_exact_1999,march_wavefunction_2000,frydel_adiabatic_2000,ludena_functional_2000,artemyev_dft_2002,amovilli_exact_2003,holas_wave_2003,march_kinetic_2003,march_effective_2004,ludena_exact_2004,katriel_study_2004,capuzzi_differential_2005,ragot_exact_2006,gomez_application_2006,zhu_exact_2006,seidl_strictly_2007,katriel_nonlocal_2007,coe_entanglement_2008,coe_erratum_2010,sun_extension_2009,gori-giorgi_study_2009,seidl_adiabatic_2010,cioslowski_robust_2015,chauhan_study_2017,kooi_local_2018,singh_semianalytical_2020,savin_second-order_2023}. 

The model has also been extended to three- to six-electron systems, primarily through the efforts of Cioslowski and coworkers \cite{taut_three_2003,cioslowski_wigner_2006,cioslowski_strong-correlation_2008,cioslowski_benchmark_2011,amovilli_hookean_2011,cioslowski_three-electron_2012,cioslowski_weak-correlation_2013,cioslowski_benchmark_2014,strasburger_order_2016,cioslowski_five-_2018}. However, the technique of variable separation is not applicable in these cases, and accurate numerical solutions are typically derived instead of analytical ones. While the many-electron version of Harmonium is an interesting topic in its own right, the original appeal of the two-electron Harmonium lies largely in its ability to reduce the coupled two-particle Hamiltonian into two uncoupled single-pseudo-particle Hamiltonians. Even when analytical solutions are unavailable for the eigenvalue problem of these single-pseudo-particle Hamiltonians, highly accurate and easily applicable numerical methods are usually available. Given this, it is reasonable to adopt the two-electron Harmonium in a way that makes it suitable for studying the electron-PCP correlation.  

\subsection{The exotic Harmonium model }

To study the electron-PCP correlation, we introduced the following Hamiltonian by replacing an electron with a PCP in the Harmonium system \cite{riyahi_quantifying_2023}:

\begin{align}
{{\hat{H}}_\mathrm{ex\textrm{-}Har}}=&{{\hat{T}}_\mathrm{e}}+{{\hat{T}}_{{\text{\tiny  PCP}}}}+\hat{V}_{\mathrm{ext}}^\mathrm{e}+\hat{V}_{\mathrm{ext}}^{{\text{\tiny  PCP}}}+{{\hat{V}}_{\mathrm{e},{\text{\tiny  PCP}}}} \nonumber \\
=& -\frac{1}{2} \nabla _\mathrm{e}^{2} -\frac{1}{2{{m}_{{\text{\tiny  PCP}}}}} \nabla _{{\text{\tiny  PCP}}}^{2} \nonumber \\
&+ \frac{{1}}{2}{\omega }^{2} r_\mathrm{e}^{2}+ \frac{{1}}{2}{{m}_{{\text{\tiny  PCP}}}}{\omega }^{2} r_{{\text{\tiny  PCP}}}^{2} \nonumber \\
&-\frac{1}{\left| {{{\br}}_{{\text{\tiny  PCP}}}}-{{{\br}}_\mathrm{e}} \right|}.
\end{align}

This model is referred to as the “exotic Harmonium”, borrowing terminology from the field of “exotic atoms” \cite{indelicato_exotic_2004,horvath_exotic_2011}. The model contains two free parameters: the mass of the PCP, $m_{{\text{\tiny  PCP}}}$, and the frequency of oscillation, $\omega$, which is shared by the two particles. As discussed in the previous section, varying these parameters allows us to simulate a wide range of real and fictitious particles, as well as diverse chemical environments. The Hamiltonian can be separated into two single-pseudo-particle problems by transforming it to the center-of-mass, $\bR=\left( {{{{\br}}_\mathrm{e}}+{{m}_{{\text{\tiny  PCP}}}}{{{\br}}_{{\text{\tiny  PCP}}}}}\right)/{M} $, and relative, $\br={{\br}_{{\text{\tiny  PCP}}}}-{{\br}_\mathrm{e}}$, coordinates:

\begin{align}
{{\hat{H}}_\mathrm{ex\textrm{-}Har}}=&{{\hat{H}}_{R}}+{{\hat{H}}_{r}}, \nonumber \\
{{\hat{H}}_{R}}=& -\frac{1}{2M} \nabla _{R}^{2}+ \frac{1}{2}M{{\omega }^{2}} {{R}^{2}}, \nonumber \\
{{\hat{H}}_{r}}=&-\frac{1}{2\mu } \nabla _{r}^{2}+ \frac{1}{2} \mu {{\omega }^{2}}{{r}^{2}}-\frac{1}{r}.
\end{align}

The center-of-mass Hamiltonian, ${{\hat{H}}_{R}}$, represents a 3D harmonic oscillator describing a pseudo-particle with a total mass, $M=1+m_{{\text{\tiny  PCP}}}$. The relative motion Hamiltonian, ${{\hat{H}}_{r}}$, describes a pseudo-particle with a reduced mass, $\mu = m_{{\text{\tiny  PCP}}}/\left( 1+m_{{\text{\tiny  PCP}}}\right)$, subject to both Coulombic attraction and harmonic potential. The ground state wavefunction of ${{\hat{H}}_{r}}$ cannot be derived analytically using the mathematical methods developed to solve the differential Schrödinger equation for the relative motion Hamiltonian of Harmonium \cite{taut_two_1993,ushveridze_quasi-exactly_1994,karwowski_harmonium_2004,karwowski_biconfluent_2014}. However, through the finite difference method we obtained an efficient numerical solution \cite{riyahi_quantifying_2023}. Let us briefly review the results gained from the previous study \cite{riyahi_quantifying_2023}.  

We employed the model to simulate hydrogen atom, $m_{{\text{\tiny  PCP}}}=1836$ and $\omega=0.01$, and the muonium atom, $m_{{\text{\tiny  PCP}}}=207$ and $\omega=0.02$, in typical protonic and muonic molecules under ambient conditions. For both systems, we computed the exact ground-state wavefunctions, ${{\Psi }_{\mathrm{exact}}}\left( {{{\br}}_\mathrm{e}},{{{\br}}_{{\text{\tiny  PCP}}}} \right)$, as well as the variationally optimized uncorrelated ground state wavefunctions, $\Psi _{\mathrm{uncorr}}^{{}}\left( {{{\br}}_\mathrm{e}},{{{\br}}_{{\text{\tiny  PCP}}}} \right)={{\phi }_\mathrm{e}}\left( {{{\br}}_\mathrm{e}} \right){{\phi }_{{\text{\tiny  PCP}}}}\left( {{{\br}}_{{\text{\tiny  PCP}}}} \right)$; for a two particle system, this is equivalent to the MC-HF wavefunction. A comparison of the exact and uncorrelated one-PCP densities, $\rho \left( {{{\br}}_{{\text{\tiny  PCP}}}} \right)=\int{d{{{\br}}_\mathrm{e}}}{{\left| \Psi \left( {{{\br}}_\mathrm{e}},{{{\br}}_{{\text{\tiny  PCP}}}} \right) \right|}^{2}}$, revealed a pathological relative “overlocalization” of the uncorrelated PCP densities; interestingly, no such overlocalization observed for the one-electron densities, $\rho \left( {{{\br}}_\mathrm{e}} \right)=\int{d{{{\br}}_{{\text{\tiny  PCP}}}}}{{\left| \Psi \left( {{{\br}}_\mathrm{e}},{{{\br}}_{{\text{\tiny  PCP}}}} \right) \right|}^{2}}$. This phenomenon aligns with observations previously reported in real protonic and muonic molecules \cite{yang_development_2017,brorsen_alternative_2018,tao_multicomponent_2019,goli_two-component_2022}. Such overlocalization in approximate solutions of the MC Schrödinger equation is interpreted as a signature of either the complete absence or improper incorporation of the electron-PCP correlation in the approximate \textit{ab initio} MC wavefunctions. A similar issue arises in the context of \textit{ab initio} MC-DFT when efficient electron-PCP correlation functionals are lacking. However, by incorporating recently designed electron-PCP correlation functionals into the MC-DFT equations, i.e., the MC Kohn-Sham (KS) equations, the overlocalization either disappears or is significantly reduced \cite{yang_development_2017,brorsen_alternative_2018,tao_multicomponent_2019,goli_two-component_2022}. This finding demonstrates that the model not only accurately simulates the pathological overlocalization but also validates the effectiveness of electron-PCP correlation functionals designed to address this issue in real molecules. This is a promising result, as the exotic Harmonium model can serve as a simple and effective tool for evaluating the quality of the future electron-PCP correlation functionals. Additionally, it was shown that none of the considered functionals could reproduce the exact correlation potential of the PCP, as deduced from the inversion of MC-KS equations, a fact that, to the best of the authors’ knowledge, had not been previously noticed. Similarly, it was observed that the exact pair correlation function and corresponding zero-distance enhancement factor cannot be derived from a recently proposed electron-muon correlation functional \cite{goli_comment_2023,deng_two-component_2023}.   
Considering these findings, we believe the exotic Harmonium model effectively captures the essential physics of the electron-PCP correlation. Due to its simplicity, the model can be extended beyond the simulation of electron-PCP correlation in hydrogen and muonium atoms. Therefore, in this study, we explore a broad range of $m_{{\text{\tiny  PCP}}}$ and $\omega$, to gain a more comprehensive understanding of the nature of electron-PCP correlation. 

\subsection{Measures and manifestations of the electron-PCP correlation}

As discussed previously, the electron-PCP correlation differs fundamentally from the electron-electron correlation due to the attractive nature of electron-PCP interactions and their mass-dependence. In this section, we develop a theoretical framework specifically tailored to address the electron-PCP correlation in the exotic Harmonium system. A broader analysis of this formalism for general many-body MC quantum systems, including its interplay with other inter-particle correlations, lies beyond the scope of the present study and will be explored in the future.

The fundamental quantity in our analysis is the pair density for two-component systems containing $N_e$ and $N_{{\text{\tiny  PCP}}}$ numbers of electrons and PCPs, respectively, defined analogously to the diagonal elements of the spinless reduced two-electron density matrix \cite{davidson_reduced_1976,mcweeny_methods_1992,coleman_reduced_2000}:
\begin{widetext}
\begin{equation}
    \Gamma \left( {{{\br}}_\mathrm{e}},{{{\br}}_{{\text{\tiny  PCP}}}} \right)={{N}_\mathrm{e}}{{N}_{{\text{\tiny  PCP}}}}\sum\limits_{spins}{{}}\int{d\br_{2}^\mathrm{e}...d\br_{{{N}_\mathrm{e}}}^\mathrm{e}d\br_{2}^{{\text{\tiny  PCP}}}...d\br_{{{N}_{{\text{\tiny  PCP}}}}}^{{\text{\tiny  PCP}}}}{{\left| \Psi \left( \br_{1}^\mathrm{e},...\br_{{{N}_\mathrm{e}}}^\mathrm{e},\br_{1}^{{\text{\tiny  PCP}}},...\br_{{{N}_{{\text{\tiny  PCP}}}}}^{{\text{\tiny  PCP}}},\left\{ spins \right\} \right) \right|}^{2}}.
\end{equation}
% \end{widetext}

This quantity represents the joint probability of simultaneously finding one electron and a PCP at two spatial locations. The pair density is decomposable into uncorrelated and correlated contributions as follows:
% \begin{widetext}
\begin{align}
    \Gamma \left( {{{\br}}_\mathrm{e}},{{{\br}}_{{\text{\tiny  PCP}}}} \right)=&\rho \left( {{{\br}}_\mathrm{e}} \right)\rho \left( {{{\br}}_{{\text{\tiny  PCP}}}} \right)+{{\Gamma }_{\mathrm{c}}}\left( {{{\br}}_\mathrm{e}},{{{\br}}_{{\text{\tiny  PCP}}}} \right), \nonumber \\
    \rho \left( {{{\br}}_\mathrm{e}} \right)=&{{N}_\mathrm{e}}\sum\limits_{spins}{{}}\int{d\br_{2}^\mathrm{e}...d\br_{{{N}_\mathrm{e}}}^\mathrm{e}d\br_{1}^{{\text{\tiny  PCP}}}...d\br_{{{N}_{{\text{\tiny  PCP}}}}}^{{\text{\tiny  PCP}}}}{{\left| \Psi \left( \br_{1}^\mathrm{e},...\br_{{{N}_\mathrm{e}}}^\mathrm{e},\br_{1}^{{\text{\tiny  PCP}}},...\br_{{{N}_{{\text{\tiny  PCP}}}}}^{{\text{\tiny  PCP}}},\left\{ spins \right\} \right) \right|}^{2}}, \nonumber \\
    \rho \left( {{{\br}}_{{\text{\tiny  PCP}}}} \right)=&{{N}_{{\text{\tiny  PCP}}}}\sum\limits_{spins}{{}}\int{d\br_{1}^\mathrm{e}...d\br_{{{N}_\mathrm{e}}}^\mathrm{e}d\br_{2}^{{\text{\tiny  PCP}}}...d\br_{{{N}_{{\text{\tiny  PCP}}}}}^{{\text{\tiny  PCP}}}}{{\left| \Psi \left( \br_{1}^\mathrm{e},...\br_{{{N}_\mathrm{e}}}^\mathrm{e},\br_{1}^{{\text{\tiny  PCP}}},...\br_{{{N}_{{\text{\tiny  PCP}}}}}^{{\text{\tiny  PCP}}},\left\{ spins \right\} \right) \right|}^{2}}.
\end{align}
\end{widetext}

In the absence of inter-component correlations, the pair density factorizes into the simple product of one-electron and one-PCP densities, mirroring the joint probability distribution of two statistically independent events. Deviations from this factorized case imply correlation and are quantified through conditional densities, which we use to define electron and PCP \textit{hills}: 

\begin{align}
    \rho _{{}}^{\mathrm{cond}}\left( {{{\br}}_\mathrm{e}};{{{\br}}_{{\text{\tiny  PCP}}}} \right)&=\frac{\Gamma \left( {{{\br}}_\mathrm{e}},{{{\br}}_{{\text{\tiny  PCP}}}} \right)}{\rho \left( {{{\br}}_{{\text{\tiny  PCP}}}} \right)} \nonumber \\
    &=\rho \left( {{{\br}}_\mathrm{e}} \right)+\rho _{\mathrm{c}}^{\mathrm{hill}}\left( {{{\br}}_\mathrm{e}};{{{\br}}_{{\text{\tiny  PCP}}}} \right), \nonumber \\
    \rho _{{}}^{\mathrm{cond}}\left( {{{\br}}_{{\text{\tiny  PCP}}}};{{{\br}}_\mathrm{e}} \right)&=\frac{\Gamma \left( {{{\br}}_{{\text{\tiny  PCP}}}},{{{\br}}_\mathrm{e}} \right)}{\rho \left( {{{\br}}_\mathrm{e}} \right)} \nonumber \\
    &=\rho \left( {{{\br}}_{{\text{\tiny  PCP}}}} \right)+\rho _{\mathrm{c}}^{\mathrm{hill}}\left( {{{\br}}_{{\text{\tiny  PCP}}}};{{{\br}}_\mathrm{e}} \right).
\end{align}

The conditional electron (PCP) density represents the probability of finding an electron (PCP) at ${{{\br}}_\mathrm{e}}$ $({\br_{{\text{\tiny  PCP}}}})$, given that a reference PCP (electron) is located at ${{{\br}}_{{\text{\tiny  PCP}}}}$ $({{{\br}}_\mathrm{e}})$. The \textit{correlation hills}, $\rho _{\mathrm{c}}^{\mathrm{hill}}\left( {{{\br}}_\mathrm{e}};{{{\br}}_{{\text{\tiny  PCP}}}} \right)={{\Gamma }_{\mathrm{c}}}\left( {{{\br}}_\mathrm{e}},{{{\br}}_{{\text{\tiny  PCP}}}} \right)/ \rho \left( {{{\br}}_{{\text{\tiny  PCP}}}} \right)$ and $\rho _{\mathrm{c}}^{\mathrm{hill}}\left( {{{\br}}_{{\text{\tiny  PCP}}}};{{{\br}}_\mathrm{e}} \right)={{\Gamma }_{\mathrm{c}}}\left( {{{\br}}_\mathrm{e}},{{{\br}}_{{\text{\tiny  PCP}}}} \right)/ \rho \left( {{{\br}}_\mathrm{e}} \right)$, quantify the deviation of the conditional densities from their corresponding single-particle densities. In the single-component electronic systems, the established concepts of \textit{Fermi holes} and \textit{Coulomb holes} quantify deviations from an uncorrelated reference state \cite{thakkar_extracules_1987,buijse_fermi_1996,baerends_quantum_1997}. These holes physically represent how each electron effectively excludes other electrons from its vicinity – a manifestation of the antisymmetric nature of electronic wavefunctions (Fermi) and the repulsive electrostatic interactions (Coulomb) between electrons \cite{kohanoff_electronic_2006}. In contrast, due to the attractive nature of electron-PCP interactions, each particle is surrounded by a correlation hill rather than a correlation hole. Furthermore, because electrons and PCPs are \textit{distinguishable} particles, no analogue of the Fermi hole arises. We emphasize that in a general two-component many-body system, the correlation picture becomes complex as both hills and holes coexist. However, in the exotic Harmonium model, this complexity simplifies dramatically, and each particle carries only a correlation hill, with no hole-like features emerging. Among the most significant properties of the correlation hills are their characteristic sum rules:

\begin{align}\label{eq:8}
    \int{d{{{\br}}_\mathrm{e}}}\rho _{\mathrm{c}}^{\mathrm{hill}}\left( {{{\br}}_\mathrm{e}};{{{\br}}_{{\text{\tiny  PCP}}}} \right)&=0, \nonumber \\
    \int{d{{{\br}}_{{\text{\tiny  PCP}}}}}\rho _{\mathrm{c}}^{\mathrm{hill}}\left( {{{\br}}_{{\text{\tiny  PCP}}}};{{{\br}}_\mathrm{e}} \right)&=0.
\end{align}

Consequently, while correlation hills create enhanced probability regions near a reference particle, they must be compensated by depleted regions at larger distances to satisfy the sum rules. Remarkably, this exact compensation property, though with the opposite sign, mirrors the behavior observed in the case of Coulomb holes \cite{thakkar_extracules_1987,buijse_fermi_1996,baerends_quantum_1997}. 

Another quantity often used to quantify correlation is the \textit{intracule density} \cite{thakkar_extracules_1987}. For the two-component systems, it is defined as follows:

\begin{equation}
    {{\rho }_{\mathrm{int}}}\left( {{{\br}}_{\mathrm{e},{\text{\tiny  PCP}}}} \right)=\int{d{{{\br}}_\mathrm{e}}\int{d{{{\br}}_{{\text{\tiny  PCP}}}}}}\Gamma \left( {{{\br}}_\mathrm{e}},{{{\br}}_{{\text{\tiny  PCP}}}} \right)\delta \left( \br-{{{\br}}_{\mathrm{e},{\text{\tiny  PCP}}}} \right).
\end{equation}

The intracule density describes the probability of finding a pair of electron-PCP separated by a specific inter-particle distance and orientation. For isotropic states, such as the ground state of the exotic Harmonium, this reduces to the radial distribution function (RDF): ${{D}_{\mathrm{int}}}\left( r \right)=4\pi {{r}^{2}}{{\rho }_{\mathrm{int}}}\left( r \right)$, $r=\left|{{{\br}}_{\mathrm{e},{\text{\tiny  PCP}}}} \right|$. Historically, the intracule RDFs have been used to define \textit{reference-dependent} Fermi holes and correlation holes in the single-component electronic systems \cite{coulson_electron_1961}: $\Delta {{D}_{\mathrm{int}}}\left( r \right)=D_{\mathrm{exact}}^{\mathrm{int}}\left( r \right)-D_{\mathrm{uncorr}}^{\mathrm{int}}\left( r \right)$. In the single-component electronic systems, the uncorrelated reference is usually the HF wavefunction. However, it might be more accurate to call this the \textit{least correlated} wavefunction, since it already includes Fermi correlation through antisymmetrization of the wavefunction. Beyond real atoms and molecules \cite{valderrama_intracule_2000}, this formalism has also proven valuable for studying the role of electron-electron correlation effects in model systems like Harmonium and other related two-electron artificial atoms \cite{samanta_correlation_1990,ghosh_study_1991,oneill_wave_2003,makarewicz_coulomb_1988,romera_electron-pair_2002}. For the two-component systems the sum rule for the intracule RDF, $\int{dr}{{D}_{\mathrm{int}}}\left( r \right)={{N}_\mathrm{e}}{{N}_{{\text{\tiny  PCP}}}}$, determines the total number of the electron-PCP pairs implying: $\int{dr}\Delta {{D}_{\mathrm{int}}}\left( r \right)=0$. In both Harmonium and Helium atoms, $\Delta {{D}_{\mathrm{int}}}\left( r \right)$ exhibits similar characteristics: a negative region at small inter-electronic distances and a positive region at larger distances \cite{oneill_wave_2003}. This behavior mirrors findings from studies of the Fermi hole and Coulomb hole, demonstrating how electron-electron correlation enhances the spatial separation of electrons. However, for the electron-PCP correlation, $\Delta {{D}_{\mathrm{int}}}\left( r \right)$ should exhibit the opposite pattern of negative and positive regions, diminishing the spatial separation of the electron-PCP pair due to their attractive interaction. 

For the exotic Harmonium, the separability of the original Hamiltonian into relative and center of mass coordinates leads to a factorizable ground state total wavefunction: $\Psi _{\mathrm{exact}}^{{}}\left( \br,\bR \right)=\psi \left( {\br} \right)\varphi \left( {\bR} \right)$. The components, $\psi$ and $\varphi$, are the ground state eigenfunctions of the relative and center-of-mass Hamiltonians, respectively. Accordingly, the exact intracule RDFs only depend on the relative coordinate eigenfunctions. The intracule RDF serves as a fundamental tool for computing several key quantities in the exotic Harmonium, including the electron-PCP interaction, $\left\langle {{{\hat{V}}}_{\mathrm{e},{\text{\tiny  PCP}}}} \right\rangle =-\int{dr\frac{D\left( r \right)}{r}}$, and statistical moments of the electron-PCP distance distribution, $\left\langle {{r}^{n}} \right\rangle =\int{dr\text{ }{{r}^{n}}}D\left( r \right)$.

Beyond their roles in determining the probabilities and spatial distribution of quantum particles, correlations may also significantly influence other observables, most notably, the energy. For single-component electronic systems, Baerends and Gritsenko's comprehensive review outlines three distinct notions of the electron-electron correlation energy \cite{baerends_quantum_1997}. While extending these concepts to general multi-component quantum systems is desirable, the present study focuses exclusively on the exotic Harmonium and the electron-PCP correlation energy, deferring the complexities of a broader generalization to a future study. Among the three definitions of the electron-PCP correlation energy \cite{baerends_quantum_1997}, one arises within the framework of MC-DFT; this has been thoroughly analyzed in our prior work \cite{riyahi_quantifying_2023}, and is not revisited here. 

Similar to the correlation hole, two distinct notions of electron-PCP correlation energy can be defined: one reference-dependent and the other \textit{reference-independent}. The \textit{reference-independent} definition arises naturally from the aforementioned pair density decomposition:

\begin{align}
    E_{\mathrm{corr}}^\mathrm{ref\textrm{-}indep}&=-\int{\int{d{{{\br}}_\mathrm{e}}d{{{\br}}_{{\text{\tiny  PCP}}}}}\frac{\rho \left( {{{\br}}_\mathrm{e}} \right)\rho _{\mathrm{c}}^{\mathrm{hill}}\left( {{{\br}}_{{\text{\tiny  PCP}}}};{{{\br}}_\mathrm{e}} \right)}{\left| {{{\br}}_{{\text{\tiny  PCP}}}}-{{{\br}}_\mathrm{e}} \right|}} \nonumber \\
    &=-\int{\int{d{{{\br}}_\mathrm{e}}d{{{\br}}_{{\text{\tiny  PCP}}}}\frac{\rho \left( {{{\br}}_{{\text{\tiny  PCP}}}} \right)\rho _{\mathrm{c}}^{\mathrm{hill}}\left( {{{\br}}_\mathrm{e}};{{{\br}}_{{\text{\tiny  PCP}}}} \right)}{\left| {{{\br}}_{{\text{\tiny  PCP}}}}-{{{\br}}_\mathrm{e}} \right|}}}.
\end{align}

The total interaction energy is the sum of the \textit{classical Coulomb} interaction energy \cite{riyahi_quantifying_2023}, $J_\mathrm{ep}$, and the correlation energy:

\begin{align}
    \left\langle {{{\hat{V}}}_{\mathrm{e},{\text{\tiny  PCP}}}} \right\rangle &={{J}_\mathrm{ep}}+E_{\mathrm{corr}}^\mathrm{ref\textrm{-}indep}, \nonumber \\
    {{J}_\mathrm{ep}}&=-\int{\int{d{{{\br}}_\mathrm{e}}d{{{\br}}_{{\text{\tiny  PCP}}}}\frac{\rho \left( {{{\br}}_\mathrm{e}} \right)\rho \left( {{{\br}}_{{\text{\tiny  PCP}}}} \right)}{\left| {{{\br}}_{{\text{\tiny  PCP}}}}-{{{\br}}_\mathrm{e}} \right|}}}.
\end{align}

However, this definition is seldom used in practice due to the lack of exact or highly accurate approximate wavefunctions for most many-body quantum systems. The exotic Harmonium is an exception where $\Psi_{\mathrm{exact}}$ is accessible, making the definition numerically applicable. 

In contrast, the reference-dependent correlation energy is defined by attributing total energy to the uncorrelated state, $\Psi _{\mathrm{uncorr}}^{{}}\left( {{{\br}}_\mathrm{e}},{{{\br}}_{{\text{\tiny  PCP}}}} \right)$, through the energy variation:
\begin{widetext}
\begin{align}
    E_{\mathrm{corr}}^\mathrm{ref\textrm{-}dep}=&{{E}_{\mathrm{exact}}}-{{E}_{\mathrm{uncorr}}}, \nonumber \\
    {{E}_{\mathrm{uncorr}}}=&\underset{\Psi _{\mathrm{uncorr}}^{{}}}{\mathop{\min }}\,\left[ \int{\int{d{{{\br}}_\mathrm{e}}d{{{\br}}_{{\text{\tiny  PCP}}}}}}\Psi _{\mathrm{uncorr}}^{{}}\left( {{{\br}}_\mathrm{e}},{{{\br}}_{{\text{\tiny  PCP}}}} \right){{{\hat{H}}}_\mathrm{ex-Har}}\Psi _{\mathrm{uncorr}}^{{}}\left( {{{\br}}_\mathrm{e}},{{{\br}}_{{\text{\tiny  PCP}}}} \right) \right].
\end{align}
\end{widetext}
This definition is equivalent to the one used widely in the one-component electronic systems: $E_{\mathrm{corr}}=E_{\mathrm{exact}}-E_{\text{\tiny HF}}$ \cite{baerends_quantum_1997}. Because the original Hamiltonian is separable into the relative and center-of-mass components, the total energy corresponds to the sum of the ground-state energies of these two Hamiltonians, $E_{\mathrm{exact}}=E_{R}+E_{r}$. The distinction from the reference-independent correlation energy becomes clearer considering that the reference-dependent correlation energy arises not only from the energy interaction term, but also from contributions of all the other terms in the Hamiltonian: 

\begin{align}
    E_{\mathrm{corr}}^\mathrm{ref\textrm{-}dep}=&\Delta {{T}_\mathrm{e}}+\Delta {{T}_{{\text{\tiny  PCP}}}}+\Delta V_{\mathrm{ext}}^\mathrm{e}+\Delta V_{\mathrm{ext}}^{{\text{\tiny  PCP}}} \nonumber \\
    &+\Delta {{V}_{\mathrm{e},{\text{\tiny  PCP}}}}.
\end{align}

Wherein, for the one-particle operators, denoted as $\hat{A}$, and the two-particle interaction operator, the differences between exact and uncorrelated expectation values are as follows: 
\begin{widetext}
\begin{align}
    \Delta A=&\int{\int{d{{{\br}}_\mathrm{e}}d{{{\br}}_{{\text{\tiny  PCP}}}}}}\Psi _{\mathrm{exact}}^{{}}\hat{A}\Psi _{\mathrm{exact}}^{{}}-\int{\int{d{{{\br}}_\mathrm{e}}d{{{\br}}_{{\text{\tiny  PCP}}}}}}\Psi _{\mathrm{uncorr}}^{{}}\hat{A}\Psi _{\mathrm{uncorr}}^{{}}, \nonumber \\
    \Delta {{V}_{\mathrm{e},{\text{\tiny  PCP}}}}=&-\int\int{\frac{{{\Gamma }_{\mathrm{exact}}}\left( {{{\br}}_\mathrm{e}},{{{\br}}_{{\text{\tiny  PCP}}}} \right)-{{\rho }_{\mathrm{uncorr}}}\left( {{{\br}}_\mathrm{e}} \right){{\rho }_{\mathrm{uncorr}}}\left( {{{\br}}_{{\text{\tiny  PCP}}}} \right)}{\left| {{{\br}}_{{\text{\tiny  PCP}}}}-{{{\br}}_\mathrm{e}} \right|}}d{{\br}_\mathrm{e}}d{{\br}_{{\text{\tiny  PCP}}}}.
\end{align}
\end{widetext}

We emphasize that, in general, the uncorrelated (or least correlated) reference state for electron-PCP correlation lacks a unique definition in multi-component systems \cite{cassam-chenai_quantum_2017}. As a result, multiple distinct reference-dependent electron-PCP correlation energies can be formulated, rather than relying solely on the definition based on the MC-HF reference state \cite{pavosevic_multicomponent_2020}. However, such ambiguity does not arise in the case of the exotic Harmonium.

\section{Results}

A broad range of values for the PCP’s mass, $1\leq m_{{\text{\tiny  PCP}}}\leq 1836$, and the vibrational frequency, $0.0001\leq \omega\leq 100$, were used in numerical calculations. However, for brevity, we report only $m_{{\text{\tiny  PCP}}}=\{1,10,207,1836\}$ as including additional mass values does not affect the main trends considered in this paper. While $m_{{\text{\tiny  PCP}}}=10$ is unphysical, the other three values (up to a negligible rounding error) correspond to the mass of positron, $m_{{\text{\tiny  PCP}}}=1$, muon, $m_{{\text{\tiny  PCP}}}=207$, and proton, $m_{{\text{\tiny  PCP}}}=1836$, spanning for small to large mass scales. For the vibrational frequencies, we report the results for $\omega=\{0.01,0.1,1,10,100\}$ in the figures, and additionally include $\omega=\{0.0001\}$ in the tables, as this subset captures the main frequency-dependent trends. The details of computational procedures used for acquiring the exact and MC-HF wavefunctions and their corresponding components of the energy expectation values are presented in Appendix.

\subsection{The nature of the electron-PCP system}

\begin{figure*}[t]
\includegraphics[width=\textwidth]{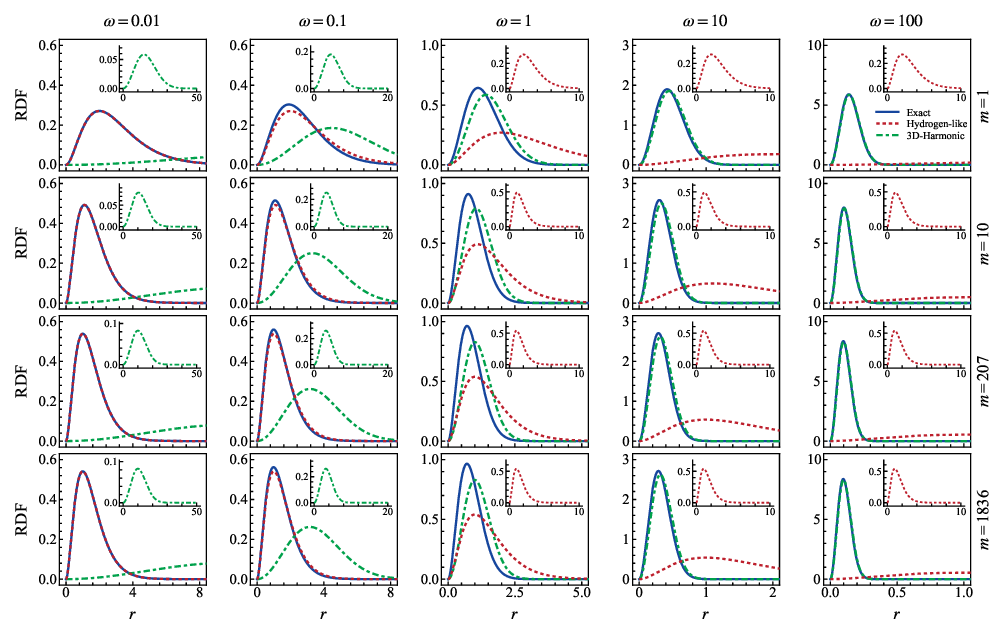}%
\caption{The exact intracule RDF of the exotic Harmonium and the corresponding functions of the hydrogen-like atom and 3D harmonic oscillator.}
\label{fig:1}
\end{figure*}

The electron-PCP system in the exotic Harmonium exhibits distinct behavior at the two extremes of vibrational frequency, $\omega \to 0$ and $\omega \to \infty$. At these limits, the relative Hamiltonian reduces to the Hamiltonians of the hydrogen-like atom, $\omega \to 0$, and the isotopic 3D harmonic oscillator, $\omega \to \infty$. To monitor the transition between these regimes at the intermediate frequencies, we examine the intracule RDF. Fig. \ref{fig:1} shows $D_{\mathrm{exact}}^{\mathrm{int}}\left( r \right)$ alongside those of the hydrogen-like atom, $D_\mathrm{hyd}^{\mathrm{int}}\left( r \right)=4{{\mu }^{3}}{{r}^{2}} e^{-2\mu r}$, and the 3D harmonic oscillator, $D_\mathrm{har}^{\mathrm{int}}\left( r \right)= 4  ({{{{\mu }^{3}}{{\omega }^{3}}}/{\pi }})^{1/2} {{r}^{2}} e^{ -2\omega {{r}^{2}} }$, for comparison.

Regardless of mass values, the transition from an atom-like system to a particle-in-trap-like system begins around $\omega=1$. This is a hundred times larger than $\omega=0.01$, which has been previously identified as the appropriate scale for simulating bonded protons and muons in molecular and condensed-phase systems under ambient conditions \cite{riyahi_quantifying_2023}. For $\omega \geq 10$, the relative system unambiguously behaves as a particle in a harmonic trap, analogous to the center-of-mass system. In the original coordinates, this corresponds to a weakly interacting electron-PCP pair, where each particle is independently confined in its own harmonic trap.

\begin{table}[t]
  \centering
  \caption{The averages and variances of the electron-PCP distance distribution.}
    \begin{ruledtabular}
    \begin{tabular}{c d{1.3}d{1.3}d{1.3}d{1.3}}
         & \multicolumn{2}{c}{$\langle r \rangle$} & \multicolumn{2}{c}{$\sigma^2_{r}$} \\
         \cmidrule{2-3}  \cmidrule{4-5}
        $\omega$   & \multicolumn{1}{c}{Exact}  & \multicolumn{1}{c}{MC-HF} & \multicolumn{1}{c}{Exact}  & \multicolumn{1}{c}{MC-HF} \\ \midrule
          & \multicolumn{4}{c}{$m=1$} \\ \cmidrule{2-5}
    0.0001 & 3.0000 & 5.9816 & 3.0000 & 7.1900 \\
    0.01  & 2.9947 & 5.8676 & 2.9811 & 6.8849 \\
    0.1   & 2.6884 & 3.8717 & 2.1219 & 2.8296 \\
    1     & 1.3501 & 1.4841 & 0.3927 & 0.3995 \\
    10    & 0.4807 & 0.4937 & 0.0437 & 0.0437 \\
    100   & 0.1572 & 0.1585 & 0.0045 & 0.0045 \\
          & \multicolumn{4}{c}{$m=10$} \\ \cmidrule{2-5}
    0.0001 & 1.6500 & 2.5908 & 0.9075 & 1.5619 \\
    0.01  & 1.6491 & 2.5791 & 0.9058 & 1.5472 \\
    0.1   & 1.5788 & 2.1296 & 0.7830 & 1.0317 \\
    1     & 0.9360 & 0.9972 & 0.2011 & 0.2020 \\
    10    & 0.3502 & 0.3552 & 0.0237 & 0.0236 \\
    100   & 0.1160 & 0.1164 & 0.0025 & 0.0025 \\
          & \multicolumn{4}{c}{$m=207$} \\ \cmidrule{2-5}
    0.0001 & 1.5072 & 1.7137 & 0.7573 & 0.8750 \\
    0.01  & 1.5066 & 1.7076 & 0.7560 & 0.8700 \\
    0.1   & 1.4512 & 1.5372 & 0.6665 & 0.7021 \\
    1     & 0.8834 & 0.8889 & 0.1812 & 0.1811 \\
    10    & 0.3336 & 0.3340 & 0.0216 & 0.0216 \\
    100   & 0.1107 & 0.1108 & 0.0022 & 0.0022 \\
          & \multicolumn{4}{c}{$m=1836$}  \\ \cmidrule{2-5}
    0.0001 & 1.5008 & 1.5704 & 0.7508 & 0.7898 \\
    0.01  & 1.5001 & 1.5624 & 0.7496 & 0.7843 \\
    0.1   & 1.4454 & 1.4594 & 0.6614 & 0.6673 \\
    1     & 0.8810 & 0.4540 & 0.1803 & 0.2901 \\
    10    & 0.3329 & 0.3329 & 0.0215 & 0.0215 \\
    100   & 0.1105 & 0.1105 & 0.0022 & 0.0022 \\
    \end{tabular}%
    \end{ruledtabular}
  \label{tab:1}%
\end{table}%

To provide a more quantitative perspective, Table \ref{tab:1} presents the average, ${{\left\langle r \right\rangle }}$, and the variance, $\sigma _{r}^{2}=\left\langle {{r}^{2}} \right\rangle -{{\left\langle r \right\rangle }^{2}}$, of the electron-PCP distance distribution, computed for both the exact and MC-HF wavefunctions. These quantities are analytically known for the hydrogen-like atoms, $\left\langle r \right\rangle ={3/(2\mu) }$, $\sigma _{r}^{2}=3/(2\mu)^{2}$, and 3D harmonic oscillator, $\left\langle r \right\rangle ={2/(\pi \mu \omega)^{1/2}}$, $\sigma _{r}^{2}=(3\pi-8)/(2\pi \mu \omega)$. The analytical predictions align well with the computed exact values at the two aforementioned extremes, corroborating the trends observed in the analysis of $D_{\mathrm{exact}}^{\mathrm{int}}\left( r \right)$. Both the exact and MC-HF results show that $\left\langle r \right\rangle $ is inversely proportional to $m_{{\text{\tiny  PCP}}}$ and $\omega$, consistent with the analytical equations. However, $\omega$ dominates this relationship and the system contracts roughly one order of magnitude by transitioning from $\omega=0.0001$ to $\omega=100$. A similar trend holds for $\sigma _{r}^{2}$, and at $\omega=100$, $\sigma _{r}^{2} < 0.005$, the distribution of electron-PCP distance reduces to a delta-like function. 

Notably, the discrepancy between the exact and MC-HF values is most pronounced in the $\omega \to 0$ limit, especially for small values of $m_{{\text{\tiny  PCP}}}$, but vanishes almost entirely at the $\omega \to \infty$ limit - regardless of $m_{{\text{\tiny  PCP}}}$. This suggests that electron-PCP correlation plays a critical role in exotic Harmonium when it exhibits the atom-like behavior, particularly for lighter atoms.

\subsection{The overlocalization of the one-PCP densities}

\begin{figure*}[t]
\includegraphics[width=\textwidth]{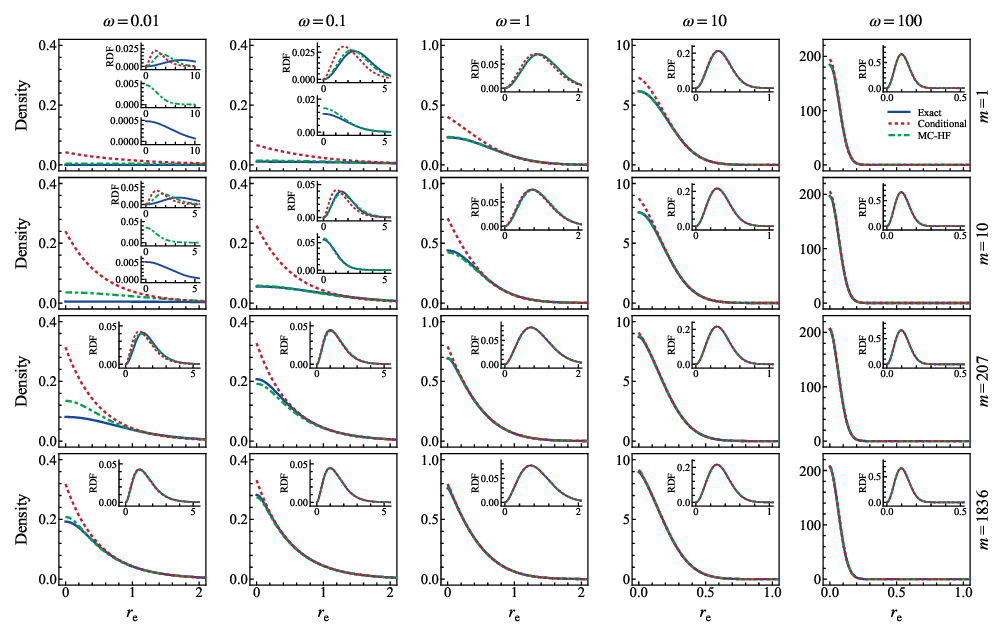}%
\caption{The 1D slices of the exact and MC-HF one- and conditional electron densities. The insets show the RDF of densities or the individual/zoomed depictions of the one-electron densities.}
\label{fig:2}
\end{figure*}

\begin{figure*}[t]
\includegraphics[width=\textwidth]{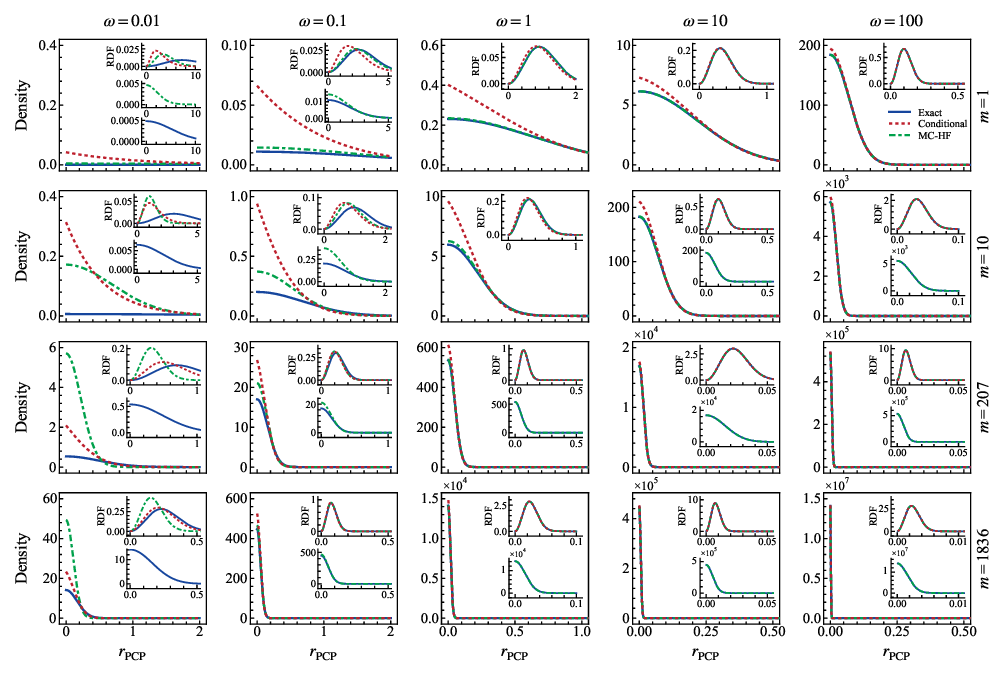}%
\caption{The 1D slices of the exact and MC-HF one- and conditional PCP densities. The insets show the RDF of densities or the individual/zoomed depictions of the one-PCP densities.}
\label{fig:3}
\end{figure*}

\begin{figure*}[t]
\includegraphics[width=\textwidth]{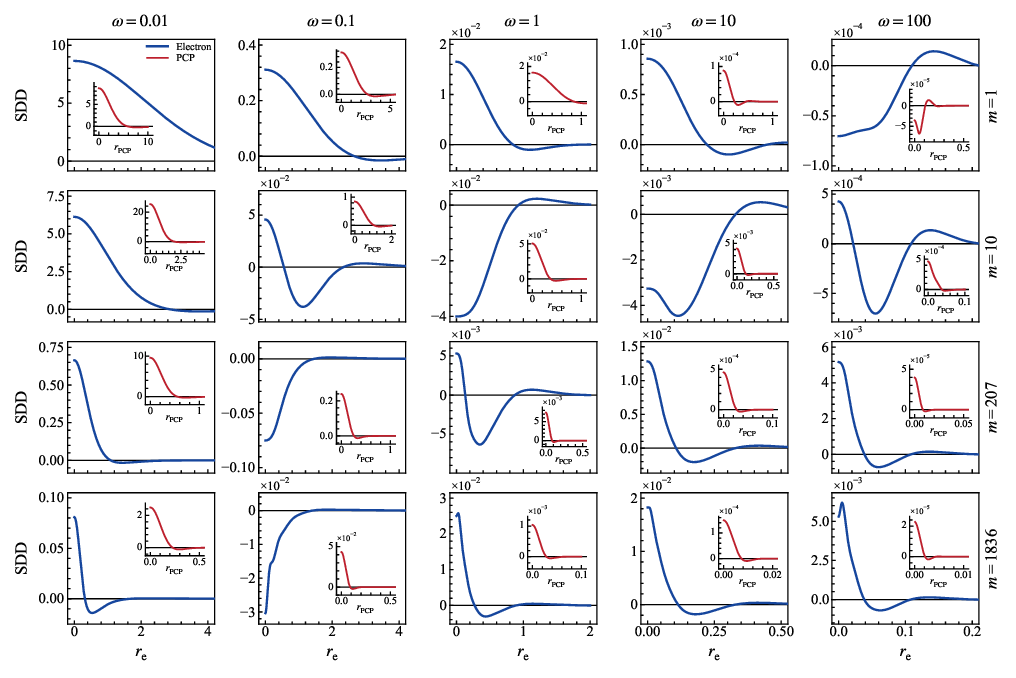}%
\caption{The scaled density difference (SDD), $(\rho_{\text{\tiny MC\textrm{-}HF}}-\rho_\mathrm{exact})/\rho_\mathrm{exact}^\mathrm{max}$, of the one-electron and one-PCP (insets) densities where $\rho_\mathrm{exact}^\mathrm{max}$ is the maximum amplitude of $\rho_\mathrm{exact}$, which appears at the origin.}
\label{fig:4}
\end{figure*}

As discussed earlier, the absence or improper inclusion of electron-PCP correlation leads to overlocalization of one-PCP densities in MC-HF and MC-DFT calculations for real molecules. Since we observed similar overlocalization in exotic Harmonium for a limited set of $m_{{\text{\tiny  PCP}}}$ and $\omega$ in our previous study \cite{riyahi_quantifying_2023}, it is essential to investigate this effect across the full parameter range considered here. Figs. \ref{fig:2} and \ref{fig:3} display the exact and MC-HF computed one-electron and one-PCP densities. Also, Fig. \ref{fig:4} displays the corresponding density differences, $\Delta \rho ={{\rho }_{\text{\tiny MC\textrm{-}HF}}}-{{\rho }_{\mathrm{exact}}}$, which for the purpose of better visibility are scaled with the maximum amplitude of ${{\rho }_{\mathrm{exact}}}$. 

The most apparent trend is that as the harmonic traps narrow with the increasing $\omega$ all computed one-particle densities become more localized. Similarly, for a fixed $\omega$, increasing $m_{{\text{\tiny  PCP}}}$ enhances the localization of the one-PCP densities, becoming a delta-like function for $\omega=100$ and $m_{{\text{\tiny  PCP}}}=1836$. For one-electron densities, a similar pattern persists, although the magnitude of localization is less severe when compared to the one-PCP densities. To obtain a more quantitative perspective, we examine the two extreme limits of vibrational frequency while the PCP is assumed to be a clamped point charge. In the $\omega \to 0$ limit, based on the previous discussions, the one-electron density approaches that of a hydrogen atom, $\rho _\mathrm{hyd}^{{}}\left( {{r}_\mathrm{e}} \right)= e^{-2{{r}_\mathrm{e}}}/{\pi }$, which agrees well with our computed density for $\omega=0.01$ and $m_{{\text{\tiny  PCP}}}=1836$. Conversely, in the $\omega \to \infty$ limit, the density converges that of a 3D harmonic oscillator, $\rho _\mathrm{har}^{{}}\left( {{r}_\mathrm{e}} \right)=(\omega / \pi )^{3/2} e^{ -\omega r_\mathrm{e}^{2} }$, matching our calculation for $\omega=100$ and $m_{{\text{\tiny  PCP}}}=1836$. Notably, between these two extremes, the one-electron densities transform in a non-trivial manner from exponential to Gaussian character across intermediate frequency ranges.   

The key finding emerges from the PCP density difference curves, which clearly demonstrate the predicted overlocalization across all ranges of $m_{{\text{\tiny  PCP}}}$ and $\omega$. This effect is most pronounced for smaller $m_{{\text{\tiny  PCP}}}$ and $\omega$ values. These results, in line with the previous observations, indicate that the electron-PCP correlation plays a particularly significant role in atom-like, strongly bonded electron-PCP pairs, becoming increasingly important as $m_{{\text{\tiny  PCP}}}$ decreases.   

\subsection{The electron-PCP correlation energies}
Indicators of the electron-PCP correlation can be classified into two broad categories: \textit{global} and \textit{local}. Global indicators describe system-wide properties without spatial dependence, where the correlation energy (the focus of this section) is the primary example. Local indicators, conversely, possess spatial resolution and can map correlation effects in real space, as will be considered in the next section. Table \ref{tab:2} presents both the reference-independent, $E_{\mathrm{corr}}^\mathrm{ref\textrm{-}indep}$, and reference-dependent, $E_{\mathrm{corr}}^\mathrm{ref\textrm{-}dep}$, electron-PCP correlation energies, along with the components of the latter, i.e., $\Delta {{T}_\mathrm{e}}$, $\Delta {{T}_{{\text{\tiny  PCP}}}}$, $\Delta V_{\mathrm{ext}}^\mathrm{e}$, $\Delta V_{\mathrm{ext}}^{{\text{\tiny  PCP}}}$, and $\Delta {{V}_{\mathrm{e},{\text{\tiny  PCP}}}}$. 

\begin{table*}[t]
  \centering
  \caption{The reference-independent and reference-dependent correlation energies, and the reference-dependent correlation components.}
  \begin{ruledtabular}
    \begin{tabular}{c d{2.3}d{2.3}d{2.3}d{2.3} d{2.3}d{2.3}d{2.3}}
         $\omega$ & \multicolumn{1}{c}{$E_\mathrm{corr}^\mathrm{ref\textrm{-}indep}$} &\multicolumn{1}{c}{$E_\mathrm{corr}^\mathrm{ref\textrm{-}dep}$} & \multicolumn{1}{c}{$\Delta T_{\mathrm{e}}$}  & \multicolumn{1}{c}{$\Delta T_{{\text{\tiny  PCP}}}$}  & \multicolumn{1}{c}{$\Delta V_\mathrm{ext}^\mathrm{e}$} & \multicolumn{1}{c}{$\Delta V_{\mathrm{ext}}^{{\text{\tiny  PCP}}}$} & \multicolumn{1}{c}{$\Delta V_{{\mathrm{e}\textrm{,}{\text{\tiny  PCP}}}}$}  \\ \midrule
          & \multicolumn{7}{c}{$m=1$}  \\ \cmidrule{2-8}
    0.0001 & -0.4887 & -0.1413 & 0.0708 & 0.0708 & 0.0000 & 0.0000 & -0.2830 \\
    0.01  & -0.3899 & -0.1283 & 0.0727 & 0.0727 & 0.0029 & 0.0029 & -0.2795 \\
    0.1   & -0.2288 & -0.0861 & 0.0570 & 0.0570 & 0.0046 & 0.0046 & -0.2095 \\
    1     & -0.1327 & -0.0601 & 0.0343 & 0.0343 & 0.0014 & 0.0014 & -0.1314 \\
    10    & -0.1084 & -0.0528 & 0.0276 & 0.0276 & 0.0004 & 0.0004 & -0.1087 \\
    100   & -0.1017 & -0.0506 & 0.0257 & 0.0257 & 0.0001 & 0.0001 & -0.1022 \\
          & \multicolumn{7}{c}{$m=10$} \\ \cmidrule{2-8}
    0.0001 & -0.8826 & -0.1973 & 0.2230 & -0.0254 & 0.0000 & 0.0001 & -0.3950 \\
    0.01  & -0.6579 & -0.1835 & 0.2226 & -0.0195 & 0.0005 & 0.0060 & -0.3930 \\
    0.1   & -0.3250 & -0.1188 & 0.1805 & -0.0052 & -0.0027 & 0.0215 & -0.3130 \\
    1     & -0.1287 & -0.0595 & 0.0842 & -0.0050 & -0.0144 & 0.0210 & -0.1454 \\
    10    & -0.0867 & -0.0432 & 0.0534 & -0.0051 & -0.0168 & 0.0184 & -0.0932 \\
    100   & -0.0765 & -0.0389 & 0.0454 & -0.0050 & -0.0170 & 0.0175 & -0.0797 \\
          & \multicolumn{7}{c}{$m=207$}  \\ \cmidrule{2-8}
    0.0001 & -0.8813 & -0.0828 & 0.1148 & -0.0318 & 0.0000 & 0.0001 & -0.1658 \\
    0.01  & -0.2886 & -0.0696 & 0.1127 & -0.0255 & 0.0000 & 0.0059 & -0.1626 \\
    0.1   & -0.0603 & -0.0266 & 0.0620 & -0.0093 & -0.0009 & 0.0096 & -0.0880 \\
    1     & -0.0237 & -0.0070 & 0.0137 & -0.0025 & -0.0022 & 0.0036 & -0.0195 \\
    10    & -0.0076 & -0.0040 & 0.0062 & -0.0014 & -0.0020 & 0.0023 & -0.0090 \\
    100   & -0.0064 & -0.0033 & 0.0047 & -0.0012 & -0.0019 & 0.0019 & -0.0069 \\
          & \multicolumn{7}{c}{$m=1836$}\\ \cmidrule{2-8}
    0.0001 & -0.6838 & -0.0332 & 0.0481 & -0.0147 & 0.0000 & 0.0001 & -0.0666 \\
    0.01  & -0.0683 & -0.0218 & 0.0437 & -0.0093 & 0.0000 & 0.0042 & -0.0604 \\
    0.1   & -0.0092 & -0.0045 & 0.0122 & -0.0019 & -0.0002 & 0.0021 & -0.0167 \\
    1     & -0.0018 & -0.0009 & 0.0019 & -0.0004 & -0.0003 & 0.0005 & -0.0026 \\
    10    & -0.0009 & -0.0005 & 0.0008 & -0.0002 & -0.0003 & 0.0003 & -0.0011 \\
    100   & -0.0008 & -0.0004 & 0.0007 & -0.0002 & -0.0004 & 0.0002 & -0.0008 \\
    \end{tabular}%
    \end{ruledtabular}
  \label{tab:2}%
\end{table*}%

The absolute values of both types of correlation energies show an inverse relationship with $m_{{\text{\tiny  PCP}}}$ and $\omega$, reaching their maximum values for the lowest considered frequency, $\omega=0.0001$. For $\omega \geq 10$ (the weakly interacting electron-PCP regime), correlation energies exhibit minimal $\omega$ dependence. Given that the total energies in this regime are substantially larger than those at lower frequencies and display strong frequency dependence (See Table \ref{tab:A2} in Appendix), we conclude that correlation effects become significantly suppressed at this range. The significant disparity between the two types of correlation energies across all the $m_{{\text{\tiny  PCP}}}$ and $\omega$ ranges, $\left| E_{\mathrm{corr}}^\mathrm{ref\textrm{-}indep} \right|>\left| E_{\mathrm{corr}}^\mathrm{ref\textrm{-}dep} \right|$, demonstrates that these are fundamentally distinct quantities, both theoretically and numerically. The difference stems from the various energy components contributing to $E_{\mathrm{corr}}^\mathrm{ref\textrm{-}dep}$. Analysis reveals that while the contributions from $\Delta V_{\mathrm{ext}}^\mathrm{e}$ and $\Delta V_{\mathrm{ext}}^{{\text{\tiny  PCP}}}$ are negligible, the remaining components play substantial roles. Although $\Delta {{V}_{\mathrm{e},{\text{\tiny  PCP}}}}$ typically dominates, the kinetic energy contributions, $\Delta {{T}_\mathrm{e}}$ and $\Delta {{T}_{{\text{\tiny  PCP}}}}$, are also significant. Consequently, $E_{\mathrm{corr}}^\mathrm{ref\textrm{-}dep}$ represents a complex interplay of multiple terms, whereas $E_{\mathrm{corr}}^\mathrm{ref\textrm{-}indep}$ derives solely from $ {{V}_{\mathrm{e},{\text{\tiny  PCP}}}}$.    

We conclude that the tightly-bonded, atom-like systems (particularly the electron-positron system) are those with the largest correlation energies. This observation aligns with the earlier discussed differences between the exact and MC-HF computed mean electron-PCP distance and also the overlocalization trends observed for the one-PCP densities.

\subsection{A local view of the electron-PCP correlation}

\begin{figure}[t]
\includegraphics[width=\columnwidth]{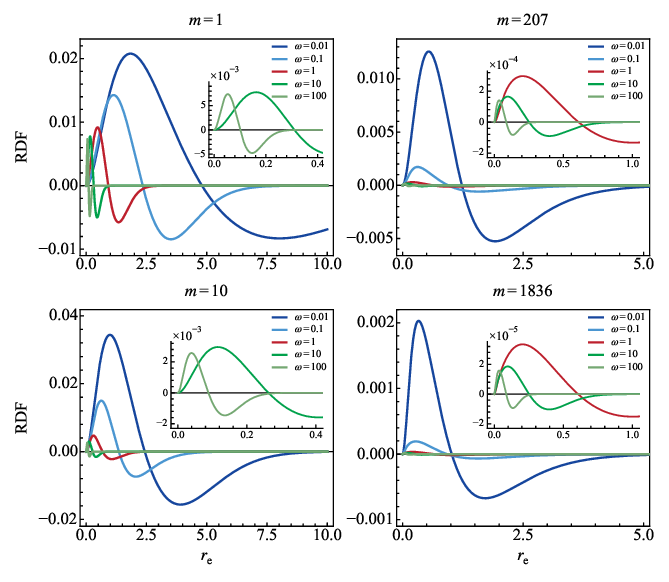}%
\caption{The RDF of the electron correlation hills where the reference PCP is placed at the origin, $\br_\mathrm{{\text{\tiny  PCP}}}=0$, i.e., the bottom of the traps. The insets are the zoomed-in views.}
\label{fig:5}
\end{figure}

\begin{figure}[t]
\includegraphics[width=\columnwidth]{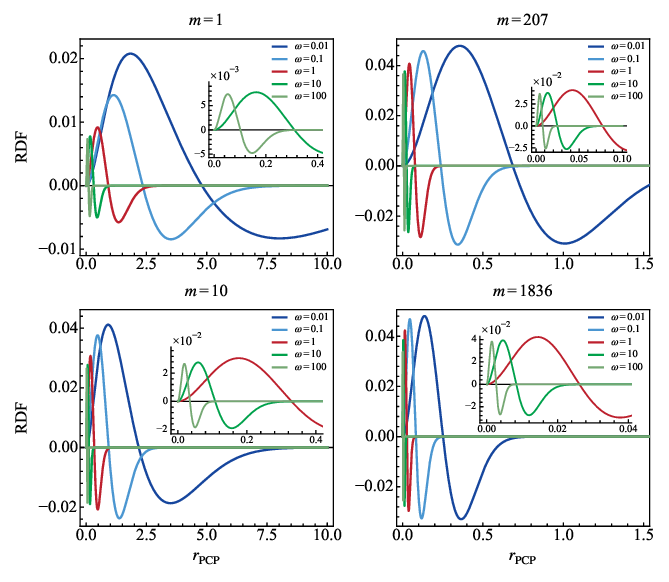}%
\caption{The RDF of the PCP correlation hills where the reference electron is placed at the origin, $\br_\mathrm{e}=0$, i.e., the bottom of the traps. The insets are the zoomed-in views.}
\label{fig:6}
\end{figure}

All global indicators of the electron-PCP correlation converge consistently, demonstrating how its relative importance varies with $m_{{\text{\tiny  PCP}}}$ and $\omega$ parameters. This raises the question: What occurs at higher values of $m_{{\text{\tiny  PCP}}}$ and $\omega$ that reduces the correlation’s significance? To address this question, we analyze local indicators, including conditional densities and correlation hills, and introduce the \textit{effective correlation radius} as a spatial measure of the electron-PCP correlation. Figs. \ref{fig:2} and \ref{fig:3} display the conditional densities of electron and PCPs, respectively, while Figs. \ref{fig:5} and \ref{fig:6} present the RDFs of the electron, $\rho _{\mathrm{c}}^{\mathrm{hill}}\left( {{{\br}}_\mathrm{e}};{{{\br}}_{{\text{\tiny  PCP}}}}=0 \right)$, and PCP, $\rho _{\mathrm{c}}^{\mathrm{hill}}\left( {{{\br}}_{{\text{\tiny  PCP}}}};{{{\br}}_\mathrm{e}}=0 \right)$, reference-independent correlation hills. 

\begin{figure}[t]
\includegraphics[width=\columnwidth]{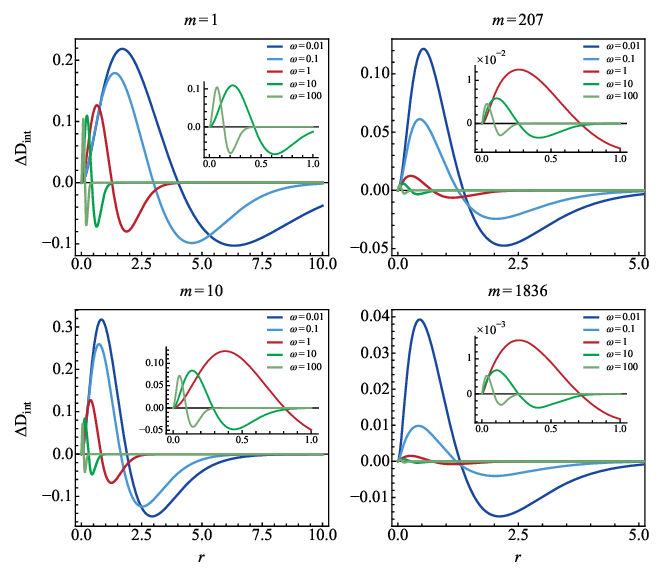}%
\caption{The difference between the exact and MC-HF intracule RDFs.}
\label{fig:7}
\end{figure}

Across the entire range of $m_{{\text{\tiny  PCP}}}$ and $\omega$, the electron and PCP conditional densities exceed their corresponding one-particle densities near the reference particle (located at the origin, i.e., the bottom of the harmonic traps). This leads to the formation of correlation hills, regions where the probability of finding the electron/PCP is enhanced compared to the uncorrelated case. Due to the sum rule (Eq. \ref{eq:8}), these hills are counterbalanced by a depletion zone at larger distances, where the probability is reduced relative to the uncorrelated scenario. This spatial pattern is the exact inverse of the behavior observed in Coulomb holes \cite{thakkar_extracules_1987,buijse_fermi_1996,baerends_quantum_1997}. Fig. \ref{fig:7}, which displays $\Delta {{D}_{\mathrm{int}}}\left( r \right)$ as the reference-dependent correlation hills, further confirms this trend: the “positive” regions consistently always appear near the origin, while the “negative” regions are shifted to larger inter-particle distances.  

To quantify spatial correlation, we determined the inter-particle distance at which the curves in Figs. \ref{fig:5} and \ref{fig:6} intersect the axis (where the correlation hills values are equal to zero: $\rho _{\mathrm{c}}^{\mathrm{hill}}\left( \br_\mathrm{e}^{c};{{{\br}}_{{\text{\tiny  PCP}}}}=0 \right)=0$ and $\rho _{\mathrm{c}}^{\mathrm{hill}}\left( \br_{{\text{\tiny  PCP}}}^{c};{{{\br}}_\mathrm{e}}=0 \right)=0$). Given the isotopic nature of these hills, we define an effective correlation sphere with radius $r_\mathrm{e}^{c}/ r_{{\text{\tiny  PCP}}}^{c}$: within this volume, the probability of finding an oppositely charged particle is enhanced. We hereafter refer to $r_\mathrm{e}^{c}/ r_{{\text{\tiny  PCP}}}^{c}$ as the effective correlation radius of electron/PCP and except for the case of $m_{{\text{\tiny  PCP}}}=1$, this radius differs for the two particles. Table \ref{tab:3} lists these radii across the full ranges of $m_{{\text{\tiny  PCP}}}$ and $\omega$.  

Both correlation radii exhibit an inverse relationship with $m_{{\text{\tiny  PCP}}}$ and $\omega$. However, they demonstrate particularly strong sensitivity to $\omega$; for a fixed $m_{{\text{\tiny  PCP}}}$, they vary up to three orders of magnitude across the studied $\omega$ range. For all $m_{{\text{\tiny  PCP}}}$ except $m_{{\text{\tiny  PCP}}}=1$, the electron’s correlation radius exceeds that of the PCP’s, and this difference becomes more pronounced at higher values of $\omega$. This observation indicates that the PCP's correlation radius is more sensitive to $\omega$ variations than that of the electron's. 

Comparison of the correlation radii with the exact $\langle r \rangle$ shows that at the $\omega \to 0$ limit, both radii exceed $\langle r \rangle$, whereas this relationship reverses in the $\omega \to \infty$ limit. Accordingly, in the $\omega \to 0$ limit, corresponding to the atom-like systems, the correlation spheres of both particles exhibit significant overlap. This substantial overlap explains the dominant role of the correlation effects in determining the global properties of these systems. The opposite behavior occurs in the $\omega \to \infty$ limit, which characterizes particle-in-trap-like systems where the correlation effects become less pronounced.  

\begin{table}[t]
  \centering
  \caption{The effective correlation radii of electrons and PCPs.}
    \begin{ruledtabular}
    \begin{tabular}{c d{2.3}d{1.3}d{1.3}d{1.3}}
        $\omega$  & \multicolumn{1}{c}{$m=1$}     & \multicolumn{1}{c}{$m=10$}    & \multicolumn{1}{c}{$m=207$}   & \multicolumn{1}{c}{$m=1836$} \\
    \midrule
          & \multicolumn{4}{c}{Electron} \\
    \cmidrule{2-5}
    0.0001 & 11.2890 & 5.7998 & 3.3411 & 2.0703  \\
    0.01  & 4.8022 & 2.4278 & 1.2299 & 1.0181   \\
    0.1   & 2.3349 & 1.3501 & 0.9419 & 0.9172   \\
    1     & 0.9109 & 0.6771 & 0.6096 & 0.6063   \\
    10    & 0.3096 & 0.2623 & 0.2513 & 0.2508   \\
    100   & 0.1001 & 0.0883 & 0.0858 & 0.0857   \\
          & \multicolumn{4}{c}{PCP} \\
    \cmidrule{2-5}
    0.0001 & 11.2890 & 5.7831 & 3.2300 & 1.7640  \\
    0.01  & 4.8022 & 2.1789 & 0.6853 & 0.2513  \\
    0.1   & 2.3349 & 0.9281 & 0.2367 & 0.0820   \\
    1     & 0.9109 & 0.3290 & 0.0771 & 0.0262   \\
    10    & 0.3096 & 0.1078 & 0.0246 & 0.0083   \\
    100   & 0.1001 & 0.0345 & 0.0078 & 0.0026   \\
    \end{tabular}%
    \end{ruledtabular}
  \label{tab:3}%
\end{table}%

\subsection{The efficiency of the adiabatic approximation}

A standard method for avoiding the direct solution of the MC Hamiltonian (Eq. \ref{eq:1}) is the \textit{adiabatic approximation}, where the MC Hamiltonian is replaced by two or more separate Hamiltonians, each describing the dynamics of a distinct type of particles \cite{baer_beyond_2006,mustroph_potential-energy_2016}. This approximation is typically applied when there is a significant mass disparity between different types of particles, i.e., electrons versus nuclei. However, recent studies have also explored its applicability, with some success, even for systems with particles of equal mass \cite{strasburger_adiabatic_2007,wolcyrz_modified_2012}. The difference between a property computed using the adiabatic approximation and that derived directly from the MC Hamiltonian is termed the \textit{non-adiabatic} contribution. Accurately calculating this contribution is computationally challenging, and for total energies, only limited data exists in the literature \cite{yang_how_2015}. This raises a key question: \textit{For MC systems, is the non-adiabatic contribution to the total energy larger than the correlation energy, or vice versa?} To our knowledge, no systematic study has addressed this question for real systems; however, exotic Harmonium offers a unique opportunity to investigate this question. 

Applying the adiabatic approximation to the Hamiltonian of exotic Harmonium yields the following two Hamiltonians for the electron and PCP, respectively: 

\begin{align}
    {{\hat{H}}_\mathrm{e}}=& -\frac{1}{2} \nabla _{{\mathrm{e}}}^{2}+ \frac{1}{2} \mu {{\omega }^{2}} r_\mathrm{e}^{2}-\frac{1}{{{r}_\mathrm{e}}}, \nonumber \\
    {{\hat{H}}_{{\text{\tiny  PCP}}}}=& -\frac{1}{2{{m}_{{\text{\tiny  PCP}}}}} \nabla _{{{{\text{\tiny  PCP}}}}}^{2}+ \frac{1}{2} {{m}_{{\text{\tiny  PCP}}}}{{\omega }^{2}} r_{{\text{\tiny  PCP}}}^{2}.
\end{align}

The total adiabatic energy is the sum of the ground-state energies of these two Hamiltonians: ${{E}_{\mathrm{ad}}}={{E}_\mathrm{e}}+{{E}_{{\text{\tiny  PCP}}}}$. ${{\hat{H}}_{{\text{\tiny  PCP}}}}$ describes an isotropic 3D harmonic oscillator, and its ground-state energy is known analytically: $E_{{\text{\tiny  PCP}}}=3\omega/2$. For ${{\hat{H}}_\mathrm{e}}$, the ground-state energy can be derived numerically using the finite difference method, analogous to the procedure discussed previously for the relative motion Hamiltonian. Table \ref{tab:4} presents the total adiabatic energies and the non-adiabatic energy contributions, ${{E}_{\textrm{non-ad}}}={{E}_{\mathrm{exact}}}-{{E}_{\mathrm{ad}}}$. For comparison, the table also offers the difference between the non-adiabatic energy contributions and the absolute values of the reference-dependent electron-PCP correlation energies: $\Delta E= {{E}_{\textrm{non-ad}}}-\left| E_{\mathrm{corr}}^\mathrm{ref\textrm{-}dep} \right|$. 

For a fixed $\omega$, $ {{E}_{\textrm{non-ad}}}$ scales inversely with $m_{{\text{\tiny  PCP}}}$ and diminishes sharply by three orders of magnitude across the considered $m_{{\text{\tiny  PCP}}}$ range. This justifies the use of the adiabatic approximation for systems carrying particles with large mass differences. Conversely, for a fixed $m_{{\text{\tiny  PCP}}}$, $ {{E}_{\textrm{non-ad}}}$ scales directly with $\omega$ and increases by one order of magnitude over the studied $\omega$ range, while, as previously discussed, $E_{\mathrm{corr}}^\mathrm{ref\textrm{-}dep}$ scales inversely with $\omega$. This explains why, for all systems except the unique case of $m_{{\text{\tiny  PCP}}}=1$, where the adiabatic approximation performs poorly, the sign of $ \Delta E$ changes at a critical $\omega$. This implies that recovering the electron-PCP correlation becomes significantly challenging, compared to $ {{E}_{\textrm{non-ad}}}$, at lower $\omega$. Interestingly, the previously determined physical frequency range under ambient conditions, $\omega \approx 0.01$, also falls within this range. Only at higher $\omega$, corresponding to extreme hydrostatic pressures, does the trend reverse. All these observations align well with the known limitations of the adiabatic approximation, beyond which the MC quantum chemical methods become indispensable. In particular, as highlighted in the introduction, positronic systems and hydrogen-rich condensed matter under high pressures emerge as prime candidates for applications of the MC quantum chemistry.

\begin{table}[t]
  \centering
  \caption{The total adiabatic energies, the non-adiabatic contributions, and the differences between the non-adiabatic contributions and the absolute values of the reference-dependent correlation energies.}
    \begin{ruledtabular}
    \begin{tabular}{c d{5.3}d{2.3}d{2.3}}
         $\omega$  & \multicolumn{1}{c}{$E_{\mathrm{ad}}$}  & \multicolumn{1}{c}{$ E_{\mathrm{non\textrm{-}ad}}$} & \multicolumn{1}{c}{$\Delta E$}   \\ \midrule
          & \multicolumn{3}{c}{$m=1$} \\ \cmidrule{2-4}
    0.0001 & -0.4999 & 0.2500 & 0.1087 \\
    0.01  & -0.4849 & 0.2501 & 0.1219 \\
    0.1   & -0.3357 & 0.2617 & 0.1756 \\
    1     & 1.6797 & 0.4322 & 0.3721 \\
    10    & 26.2654 & 1.1299 & 1.0771 \\
    100   & 288.5572 & 3.3849 & 3.3343 \\
          & \multicolumn{3}{c}{$m=10$} \\ \cmidrule{2-4}
    0.0001 & -0.4999 & 0.0455 & -0.1519 \\
    0.01  & -0.4849 & 0.0455 & -0.1380 \\
    0.1   & -0.3357 & 0.0468 & -0.0720 \\
    1     & 1.6797 & 0.0716 & 0.0121 \\
    10    & 26.2654 & 0.1816 & 0.1384 \\
    100   & 288.5572 & 0.5397 & 0.5008 \\
          & \multicolumn{3}{c}{$m=207$} \\ \cmidrule{2-4}
    0.0001 & -0.4999 & 0.0024 & -0.0804 \\
    0.01  & -0.4849 & 0.0024 & -0.0672 \\
    0.1   & -0.3357 & 0.0025 & -0.0241 \\
    1     & 1.6797 & 0.0037 & -0.0032 \\
    10    & 26.2654 & 0.0094 & 0.0054 \\
    100   & 288.5572 & 0.0279 & 0.0246 \\
          & \multicolumn{3}{c}{$m=1836$}  \\ \cmidrule{2-4}
    0.0001 & -0.4999 & 0.0003 & -0.0329 \\
    0.01  & -0.4849 & 0.0003 & -0.0216 \\
    0.1   & -0.3357 & 0.0003 & -0.0042 \\
    1     & 1.6797 & 0.0004 & -0.0005 \\
    10    & 26.2654 & 0.0011 & 0.0006 \\
    100   & 288.5572 & 0.0032 & 0.0027 \\
    \end{tabular}%
    \end{ruledtabular}
  \label{tab:4}%
\end{table}%

\section{Conclusions and prospects}

In our previous study \cite{riyahi_quantifying_2023}, we characterized the original Harmonium model as an ideal theoretical “laboratory” for investigating various aspects of the electron-electron correlation. The present work demonstrates that this is also true for the exotic Harmonium model, providing valuable insights into the electron-PCP correlation. These studies, however, merely reveal the tip of an iceberg; the model remains ripe for further exploration and extensions into new domains. Let us briefly discuss some possible extensions, applications and theoretical implications of the model.    

While previous studies explored limited regions of the system’s \textit{parameter space}, i.e., $m_{{\text{\tiny  PCP}}}$ and $\omega$ \cite{riyahi_quantifying_2023,goli_comment_2023}, the present work investigates a significantly broader range. However, as a purely numerical study, this study does not elucidate the \textit{analytical} dependence of computed quantities on the system parameters. To derive such relationships, two complementary approaches present themselves. The first approach is based on developing ultra-compact, parameter-dependent variational wavefunctions that preserve both accuracy and simplicity, akin to those recently constructed for few-electron atomic systems \cite{turbiner_ultra-compact_2021}. Such wavefunctions would enable analytical insights while maintaining computational tractability. The second approach builds on earlier work on the standard Harmonium model \cite{ghosh_study_1991,turbiner_two_1994,samanta_density-functional_1991}, where slight modifications to the electron-electron interaction transformed the system from a quasi-exactly to an exactly solvable model. An analogous strategy could be applied to the exotic Harmonium case to yield exact solutions.    

From a border perspective, a general two-component system comprising electrons and a distinct class of PCPs exhibits three correlation types: electron-electron, electron-PCP, and PCP-PCP. As emphasized in the introduction, the interdependence among these three correlations remains actively debated in the literature and is worthy of further studies \cite{brorsen_alternative_2018,sirjoosingh_multicomponent_2012,udagawa_electron-nucleus_2014}. The simplest system encompassing all three correlation types would be an extended exotic Harmonium containing two electrons and two PCPs in the harmonic confinement. Alternatively, three particle systems (either two electrons and a PCP or two PCPs and one electron) can be constructed to isolate just two out of the three correlations. These extensions naturally parallel the many-electron generalization of the standard Harmonium model previously investigated \cite{taut_three_2003,cioslowski_wigner_2006,cioslowski_strong-correlation_2008,cioslowski_benchmark_2011,amovilli_hookean_2011,cioslowski_three-electron_2012,cioslowski_weak-correlation_2013,cioslowski_benchmark_2014,strasburger_order_2016,cioslowski_five-_2018}. 

Beyond its utility for studying the electron-PCP correlation, the exotic Harmonium model may find additional applications when extended to many-particle systems. The general Hamiltonian of such a many-particle system is given by: 

\begin{align}
    {{\hat{H}}}=& -\frac{1}{2} \sum\limits_{i}^{{{N}_\mathrm{e}}}{\nabla _{\mathrm{e},i}^{2}}- \frac{1}{2{{m}_{{\text{\tiny  PCP}}}}} \sum\limits_{j}^{{{N}_{{\text{\tiny  PCP}}}}}{\nabla _{{\text{\tiny  PCP}},j}^{2}} \nonumber \\
    &+ \frac{1}{2}{{k}_\mathrm{e}}\sum\limits_{i}^{{{N}_\mathrm{e}}}{r_{\mathrm{e},i}^{2}}+ \frac{1}{2}{{k}_{{\text{\tiny  PCP}}}} \sum\limits_{j}^{{{N}_{{\text{\tiny  PCP}}}}}{r_{{\text{\tiny  PCP}},j}^{2}} \nonumber \\
    &-\sum\limits_{i}^{{{N}_\mathrm{e}}}{\sum\limits_{j}^{{{N}_{{\text{\tiny  PCP}}}}}{\frac{1}{\left| {{{\br}}_{\mathrm{e},i}}-{{{\br}}_{{\text{\tiny  PCP}},j}} \right|}}}
\end{align}

For the two-particle, the electron-PCP case, the model recovers the exotic Harmonium Hamiltonian when: $N_\mathrm{e}=N_{{\text{\tiny  PCP}}}=1$ and ${k_{{\text{\tiny  PCP}}}}/{k_\mathrm{e}}=m_{{\text{\tiny  PCP}}}$. Another special case, $N_\mathrm{e}=N_{{\text{\tiny  PCP}}}=1$ and $k_\mathrm{e}=0$, has been recently investigated as well \cite{stetzler_comparison_2023}. The model also encompasses the above-mentioned three- and four-particle systems by properly adjusting $N_\mathrm{e}$, $N_{{\text{\tiny  PCP}}}$, $k_\mathrm{e}$ and $k_{{\text{\tiny  PCP}}}$. But beyond these few-particle systems, the thermodynamics limit of the charge-neutral ($N_\mathrm{e}=N_{{\text{\tiny  PCP}}}$) version of the extended model in a single trap: $\lim N_\mathrm{e} \to \infty$, $\lim N_{{\text{\tiny  PCP}}} \to \infty$ and $k_{{\text{\tiny  PCP}}}=k_\mathrm{e}=k$, merits particular attention. For $m_{{\text{\tiny  PCP}}}=m_\mathrm{proton}$ or $m_{{\text{\tiny  PCP}}}=m_\mathrm{deuterium}$, such a system effectively simulates condensed hydrogen systems under external pressure if $k$ is gauged properly with external hydrostatic pressure. Alternatively, for $m_{{\text{\tiny  PCP}}}=m_\mathrm{positron}$, a system composed of positronium atoms in a trap is modeled, which has been in focus recently in the low-energy positron physics community \cite{mills_physics_2011}. Notably, at the thermodynamic limit, the charge-neutral system may exhibit quantum phase transitions as a function of $m_{{\text{\tiny  PCP}}}$ and $k$ \cite{sachdev_quantum_2011}, a fascinating possibility that warrants further investigation.  

From a theoretical perspective, the exotic Harmonium model could play a pivotal role in designing a future generation of efficient electron-PCP correlation functionals to be used in MC-KS equations. Specifically, analytical access to the electron and PCP correlation potentials in this system may enable the development of versatile correlation functionals applicable to a broad range of species, including particles of varying masses, under both ambient and high-pressure situations. These directions are now being actively pursued in our laboratory.

% \section*{Supplementary material}
% The supplementary material contains ...

\begin{acknowledgments}
Computational resources are provided by the SARMAD cluster.
\end{acknowledgments}

\appendix*
\section{Computational methods and data}

\begingroup
\squeezetable
\begin{table*}[t]
  \centering
  \caption{The optimized exponents of the electron and PCP (in parentheses) uncontracted [7s:7s] Gaussian basis functions for the MC-HF wavefunctions.}
    \begin{ruledtabular}
    \begin{tabular}{c  S[table-format=4.3(3.3), parse-numbers=false] S[table-format=4.3(3.3), parse-numbers=false] S[table-format=4.3(3.3), parse-numbers=false] S[table-format=4.3(3.3), parse-numbers=false] S[table-format=5.3(4.3), parse-numbers=false] S[table-format=5.3(4.3), parse-numbers=false] S[table-format=6.3(4.3), parse-numbers=false]}
        $\omega$  & \multicolumn{7}{c}{Exponent} \\ \midrule %& \multicolumn{1}{c}{2s} & \multicolumn{1}{c}{3s} & \multicolumn{1}{c}{4s} & \multicolumn{1}{c}{5s} & \multicolumn{1}{c}{6s} & \multicolumn{1}{c}{7s} \\ \midrule
          & \multicolumn{7}{c}{$m=1$}  \\   \cmidrule{2-8}

        0.0001 & 0.007 \,(0.015) & 0.010 \,(0.029) & 0.014 \,(0.046) & 0.024 \,(0.065) & 0.032 \,(0.106) & 0.056 \,(0.160) & 0.110 \,(0.195) \\
    0.01  & 0.012 \,(0.012) & 0.021 \,(0.015) & 0.033 \,(0.030) & 0.050 \,(0.062) & 0.093 \,(0.153) & 0.208 \,(0.190) & 0.270 \,(0.513) \\
    0.1   & 0.053 \,(0.044) & 0.074 \,(0.070) & 0.121 \,(0.136) & 0.155 \,(0.297) & 0.422 \,(5.940) & 0.515 \,(8.404) & 0.664 \,(11.981) \\
    1     & 0.465 \,(0.387) & 0.542 \,(0.500) & 0.764 \,(0.641) & 1.004 \,(1.107) & 1.932 \,(1.266) & 3.817 \,(1.423) & 5.420 \,(1.741) \\
    10    & 4.765 \,(3.047) & 5.125 \,(3.850) & 7.982 \,(5.072) & 19.734 \,(7.734) & 36.104 \,(8.400) & 59.889 \,(10.838) & 73.122 \,(13.891) \\
    100   & 39.372 \,(48.750) & 44.683 \,(52.497) & 51.760 \,(81.117) & 127.57 \,(120.72) & 165.38 \,(122.03) & 346.08 \,(167.20) & 402.56 \,(204.92) \\
          & \multicolumn{7}{c}{$m=10$}  \\   \cmidrule{2-8}
    0.0001 & 0.023 \,(0.195) & 0.025 \,(0.320) & 0.044 \,(0.535) & 0.087 \,(1.046) & 0.176 \,(1.377) & 0.371 \,(4.404) & 0.806 \,(6.356) \\
    0.01  & 0.035 \,(0.291) & 0.045 \,(0.532) & 0.066 \,(1.261) & 0.132 \,(1.640) & 0.227 \,(5.440) & 0.414 \,(7.404) & 0.868 \,(11.856) \\
    0.1   & 0.046 \,(0.570) & 0.073 \,(0.796) & 0.078 \,(1.035) & 0.144 \,(1.140) & 0.289 \,(5.940) & 0.614 \,(8.904) & 1.328 \,(11.981) \\
    1     & 0.441 \,(4.438) & 0.569 \,(5.375) & 1.004 \,(10.137) & 2.153 \,(20.080) & 4.763 \,(136.90) & 7.338 \,(199.04) & 13.339 \,(292.40) \\
    10    & 4.960 \,(37.400) & 5.125 \,(45.900) & 8.104 \,(51.000) & 18.269 \,(118.67) & 47.628 \,(172.60) & 251.25 \,(425.78) & 567.70 \,(467.50) \\
    100   & 37.907 \,(321.25) & 40.289 \,(486.25) & 51.760 \,(611.90) & 120.98 \,(949.00) & 316.30 \,(1359.0) & 816.63 \,(1977.8) & 1702.5 \,(2902.4) \\
          & \multicolumn{7}{c}{$m=207$}   \\ \cmidrule{2-8}
    0.0001 & 0.056 \,(2.035) & 0.128 \,(3.070) & 0.291 \,(4.140) & 0.667 \,(5.796) & 1.523 \,(8.440) & 3.337 \,(11.404) & 6.866 \,(13.856) \\
    0.01  & 0.057 \,(1.035) & 0.129 \,(2.070) & 0.291 \,(4.140) & 0.663 \,(5.796) & 1.512 \,(6.440) & 3.333 \,(11.404) & 6.958 \,(15.856) \\
    0.1   & 0.078 \,(1.035) & 0.144 \,(2.070) & 0.289 \,(4.140) & 0.614 \,(5.796) & 1.422 \,(8.940) & 3.698 \,(11.404) & 10.765 \,(14.481) \\
    1     & 0.441 \,(25.875) & 0.569 \,(56.750) & 1.004 \,(103.51) & 2.255 \,(124.08) & 5.838 \,(178.40) & 17.339 \,(281.04) & 61.670 \,(388.90) \\
    10    & 5.125 \,(113.50) & 5.253 \,(237.40) & 8.104 \,(258.40) & 18.269 \,(268.67) & 48.800 \,(472.60) & 151.25 \,(1017.5) & 567.70 \,(1038.3) \\
    100   & 37.419 \,(258.75) & 38.824 \,(517.50) & 51.760 \,(1035.0) & 123.66 \,(1449.0) & 349.99 \,(2484.0) & 1152.6 \,(10086.0) & 4702.5 \,(10353.0) \\
          & \multicolumn{7}{c}{$m=1836$} \\  \cmidrule{2-8}
    0.0001 & 0.057 \,(2.035) & 0.129 \,(3.070) & 0.291 \,(4.140) & 0.667 \,(5.796) & 1.624 \,(8.940) & 4.485 \,(12.404) & 14.304 \,(20.856) \\
    0.01  & 0.057 \,(1.035) & 0.129 \,(2.070) & 0.289 \,(4.140) & 0.669 \,(5.796) & 1.656 \,(6.940) & 4.665 \,(13.654) & 15.364 \,(22.544) \\
    0.1   & 0.070 \,(1.035) & 0.135 \,(2.070) & 0.306 \,(4.140) & 0.782 \,(5.796) & 2.240 \,(8.940) & 7.488 \,(11.404) & 32.452 \,(94.481) \\
    1     & 0.560 \,(25.875) & 1.072 \,(56.750) & 2.587 \,(103.51) & 6.416 \,(174.08) & 18.108 \,(328.40) & 64.259 \,(913.90) & 291.94 \,(931.04) \\
    10    & 5.122 \,(113.50) & 8.295 \,(268.67) & 18.299 \,(862.40) & 48.678 \,(1017.5) & 157.11 \,(6508.4) & 605.79 \,(8722.6) & 2864.6 \,(9288.3) \\
    100   & 31.622 \,(258.75) & 39.372 \,(517.50) & 51.775 \,(1035.0) & 133.19 \,(1449.0) & 421.28 \,(79047.0) & 1634.0 \,(90086.0) & 9241.6 \,(92853.0) \\

    \end{tabular}%
    \end{ruledtabular}
  \label{tab:A1}%
\end{table*}%
\endgroup

\begingroup
\squeezetable
\begin{table*}[t]
  \centering
  \caption{The total energies and expectation values of energy components derived from both exact and MC-HF (in parentheses) wavefunctions.}
    \begin{ruledtabular}
    \begin{tabular}{c  S[table-format=5.3(5.5), parse-numbers=false] S[table-format=4.3(4.5), parse-numbers=false] S[table-format=4.3(4.5), parse-numbers=false] S[table-format=4.3(4.5), parse-numbers=false] S[table-format=5.3(4.5), parse-numbers=false] S[table-format=5.3(4.5), parse-numbers=false]}
        $\omega$  & {$E$} & \multicolumn{1}{c}{$T_{\mathrm{e}}$} & \multicolumn{1}{c}{$T_{{{\text{\tiny  PCP}}}}$} & \multicolumn{1}{c}{$V_{\mathrm{ext}}^{\mathrm{e}}$} & \multicolumn{1}{c}{$V_{\mathrm{ext}}^{{{\text{\tiny  PCP}}}}$} & \multicolumn{1}{c}{$V_{{\mathrm{e}\textrm{,}{\text{\tiny  PCP}}}}$} \\ \midrule
          & \multicolumn{6}{c}{$m=1$}  \\ \cmidrule{2-7}
    0.0001 & -0.2499 \,(-0.1085) & 0.1250 \,(0.0543) & 0.1250 \,(0.0543) & 0.0000 \,(0.0000) & 0.0000 \,(0.0000) & -0.5000 \,(-0.2170) \\
    0.01  & -0.2347 \,(-0.1064) & 0.1290 \,(0.0563) & 0.1290 \,(0.0563) & 0.0039 \,(0.0010) & 0.0039 \,(0.0010) & -0.5006 \,(-0.2211) \\
    0.1   & -0.0739 \,(0.0122) & 0.1846 \,(0.1275) & 0.1846 \,(0.1275) & 0.0492 \,(0.0445) & 0.0492 \,(0.0445) & -0.5414 \,(-0.3320) \\
    1     & 2.1119 \,(2.1719) & 0.8999 \,(0.8656) & 0.8999 \,(0.8656) & 0.6519 \,(0.6505) & 0.6519 \,(0.6505) & -0.9918 \,(-0.8604) \\
    10    & 27.3953 \,(27.4481) & 7.8576 \,(7.8300) & 7.8576 \,(7.8300) & 7.1851 \,(7.1847) & 7.1851 \,(7.1847) & -2.6900 \,(-2.5813) \\
    100   & 291.9421 \,(291.9928) & 76.0373 \,(76.0116) & 76.0373 \,(76.0116) & 74.0028 \,(74.0027) & 74.0028 \,(74.0027) & -8.1380 \,(-8.0358) \\
          & \multicolumn{6}{c}{$m=10$} \\ \cmidrule{2-7}
    0.0001 & -0.4544 \,(-0.2570) & 0.4132 \,(0.1902) & 0.0414 \,(0.0668) & 0.0000 \,(0.0000) & 0.0001 \,(0.0000) & -0.9091 \,(-0.5141) \\
    0.01  & -0.4394 \,(-0.2559) & 0.4142 \,(0.1917) & 0.0482 \,(0.0677) & 0.0008 \,(0.0003) & 0.0068 \,(0.0008) & -0.9094 \,(-0.5164) \\
    0.1   & -0.2889 \,(-0.1701) & 0.4465 \,(0.2659) & 0.1121 \,(0.1173) & 0.0204 \,(0.0230) & 0.0695 \,(0.0480) & -0.9374 \,(-0.6244) \\
    1     & 1.7513 \,(1.8109) & 1.1751 \,(1.0909) & 0.7925 \,(0.7975) & 0.5133 \,(0.5277) & 0.7263 \,(0.7054) & -1.4559 \,(-1.3105) \\
    10    & 26.4471 \,(26.4904) & 8.4167 \,(8.3633) & 7.5917 \,(7.5968) & 6.7289 \,(6.7457) & 7.4229 \,(7.4045) & -3.7131 \,(-3.6199) \\
    100   & 289.0969 \,(289.1358) & 77.5784 \,(77.5330) & 75.2578 \,(75.2628) & 72.5555 \,(72.5725) & 74.7555 \,(74.7381) & -11.0503 \,(-10.9706) \\
          & \multicolumn{6}{c}{$m=207$}  \\ \cmidrule{2-7}
    0.0001 & -0.4974 \,(-0.4147) & 0.4952 \,(0.3804) & 0.0025 \,(0.0343) & 0.0000 \,(0.0000) & 0.0001 \,(0.0000) & -0.9952 \,(-0.8294) \\
    0.01  & -0.4824 \,(-0.4129) & 0.4955 \,(0.3829) & 0.0099 \,(0.0354) & 0.0002 \,(0.0002) & 0.0075 \,(0.0016) & -0.9955 \,(-0.8329) \\
    0.1   & -0.3332 \,(-0.3066) & 0.5225 \,(0.4605) & 0.0772 \,(0.0864) & 0.0141 \,(0.0150) & 0.0747 \,(0.0651) & -1.0216 \,(-0.9336) \\
    1     & 1.6834 \,(1.6904) & 1.2497 \,(1.2360) & 0.7524 \,(0.7549) & 0.4798 \,(0.4820) & 0.7487 \,(0.7451) & -1.5472 \,(-1.5277) \\
    10    & 26.2749 \,(26.2789) & 8.5584 \,(8.5523) & 7.5051 \,(7.5065) & 6.6171 \,(6.6191) & 7.4957 \,(7.4935) & -3.9015 \,(-3.8925) \\
    100   & 288.5852 \,(288.5885) & 77.9605 \,(77.9558) & 75.0143 \,(75.0155) & 72.2002 \,(72.2021) & 74.9865 \,(74.9845) & -11.5763 \,(-11.5694) \\
          & \multicolumn{6}{c}{$m=1836$}\\ \cmidrule{2-7}
    0.0001 & -0.4996 \,(-0.4664) & 0.4994 \,(0.4514) & 0.0003 \,(0.0150) & 0.0000 \,(0.0000) & 0.0001 \,(0.0000) & -0.9994 \,(-0.9328) \\
    0.01  & -0.4846 \,(-0.4628) & 0.4998 \,(0.4561) & 0.0078 \,(0.0171) & 0.0002 \,(0.0002) & 0.0075 \,(0.0033) & -0.9998 \,(-0.9393) \\
    0.1   & -0.3354 \,(-0.3309) & 0.5264 \,(0.5142) & 0.0752 \,(0.0772) & 0.0138 \,(0.0139) & 0.0750 \,(0.0729) & -1.0258 \,(-1.0091) \\
    1     & 1.6801 \,(1.6810) & 1.2535 \,(1.2516) & 0.7503 \,(0.7507) & 0.4781 \,(0.4784) & 0.7499 \,(0.7493) & -1.5517 \,(-1.5490) \\
    10    & 26.2665 \,(26.2670) & 8.5657 \,(8.5649) & 7.5006 \,(7.5008) & 6.6114 \,(6.6117) & 7.4995 \,(7.4992) & -3.9106 \,(-3.9095) \\
    100   & 288.5604 \,(288.5608) & 77.9800 \,(77.9793) & 75.0016 \,(75.0018) & 72.1820 \,(72.1824) & 74.9985 \,(74.9982) & -11.6017 \,(-11.6010) \\
    \end{tabular}%
    \end{ruledtabular}
  \label{tab:A2}%
\end{table*}%
\endgroup

For the numerical analysis, we solved the relative motion Hamiltonian eigenvalue problem using the finite difference method, implemented via a previously developed and tested computer code designed in our group to solve the one-particle Schrödinger equation \cite{riyahi_quantifying_2023}. In this method, the problem domain is discretized into finite elements, and the differential operator is transformed into a generalized matrix eigenvalue problem. This allows for efficient numerical computation of eigenvalues and eigenfunctions \cite{brenner_mathematical_2008}. To ensure the convergence of the computed quantities, we employed various grid sizes and boundary conditions. Consequently, all tabulated values are well-converged and readily reproducible by a similar code. In the first step, this approach yields the exact ground-state wavefunction, $\Psi _{\mathrm{exact}}^{{}}\left( \br,\bR \right)$, and energy, $E_{r}$. Subsequently, we computed additional quantities using this exact wavefunction. When necessary, quantities in the original variables were obtained by applying a back transformation (details discussed elsewhere \cite{laufer_test_1986}), to derive the exact wavefunction in the original coordinates, ${{\Psi }_{\mathrm{exact}}}\left( {{{\br}}_\mathrm{e}},{{{\br}}_{{\text{\tiny  PCP}}}} \right)$, which was used to calculate the desired properties. 

To compute the uncorrelated wavefunction, we employed a modified version of our group’s existing MC-HF calculation code \cite{riyahi_quantifying_2023,rayka_toward_2018}. The MC-HF calculations yield two primary results: the variationally optimized wavefunction, $\Psi _{\mathrm{uncorr}}^{{}}\left( {{{\br}}_\mathrm{e}},{{{\br}}_{{\text{\tiny  PCP}}}} \right)$, and the total energy, $E_{\mathrm{uncorr}}$. From this wavefunction, we can subsequently compute any other relevant quantity. In practice, both electronic, ${{\phi }_\mathrm{e}}$, and PCP, ${{\phi }_{{\text{\tiny  PCP}}}}$, orbitals were expanded using uncontracted Gaussian functions with exponents fully optimized for each combination of $m_{{\text{\tiny  PCP}}}$ and $\omega$. Through systematic testing, we determined that a [7s:7s] basis set provides sufficient accuracy across the entire parameter range. The inclusion of additional s-type functions or higher angular momentum functions, p- and d-type, was found to have a negligible impact on the reported results. The final optimized exponents of this basis set are provided in Table \ref{tab:A1}. The only exceptions are in the case of the MC-HF intracule densities in Fig. \ref{fig:7}, for which to remedy the numerical instabilities, the nuclear basis set was reduced to a single s-type Gaussian function while the electronic basis set was expanded up to 24 s-type functions. Also, Table \ref{tab:A2} presents the total energies and the expectation values of energy terms derived from both exact and MC-HF wavefunctions. 

% \section*{AUTHOR DECLARATIONS}
% \subsection*{Conflict of Interest}
% The authors have no conflicts to disclose.

% \section*{Data Availability}
% The data that support the findings of this study are available within the article and its supplementary material.

% Reference section
% \bibliography{references.bib}


%

\end{document}